\documentclass[twocolumn]{aastex631}

\usepackage[utf8]{inputenc}
\DeclareUnicodeCharacter{2212}{-}
\usepackage{graphicx}
\graphicspath{{./}{./figures/}{./figures/}{./membershipfigures/}}
\usepackage{amsmath}
\usepackage{afterpage}
\usepackage{comment}
\usepackage{xcolor}
\usepackage{float}
\usepackage{enumitem}
\usepackage{placeins}

\setenumerate{itemsep=0mm}

\usepackage{soul} 
\usepackage{amsmath}
\usepackage{amssymb}
\usepackage{xspace}
\usepackage{xifthen}
\usepackage{eso-pic}

\newcommand{\Gaia}{{\it Gaia}\xspace}

\definecolor{forestgreen}{HTML}{228B22}
\definecolor{urlblue}{HTML}{000000}

\mathchardef\mhyphen="2D

\newcommand{\roughly}{\ensuremath{ {\sim}\,} }
\newcommand{\gtr}{\ensuremath{ {>}\,} }

\newlength{\dhatheight}

\newcommand{\code}[1]{\texttt{#1}\xspace}

\newcommand{\unit}[1]{\ensuremath{\mathrm{\,#1}}\xspace}

\newcommand{\Gyr}{\unit{Gyr}}

\newcommand{\degree}{\ensuremath{{}^{\circ}}\xspace}

\newcommand{\km}{\unit{km}}
\newcommand{\kms}{\km \second^{-1}}

\newcommand{\second}{\unit{s}}

\newcommand{\e}{\unit{e^{-}}}

\newcommand{\secref}[1]{Section~\ref{sec:#1}}
\newcommand{\appref}[1]{Appendix~\ref{app:#1}}
\newcommand{\tabref}[1]{Table~\ref{tab:#1}}

\newcommand{\figref}[1]{Figure~\ref{fig:#1}}

\newcommand{\bandvar}[2][]{%
  \ifthenelse{\isempty{#1}}{\var{#2}}{\var{#2\_#1}}%
}

\newcommand{\ra}{{\ensuremath{\alpha_{2000}}}\xspace}
\newcommand{\dec}{{\ensuremath{\delta_{2000}}}\xspace}
\newcommand{\age}{{\ensuremath{\tau}}\xspace}

\newcommand{\PA}{\ensuremath{\mathrm{P.A.}}\xspace}

\newcommand{\HEALPix}{\code{HEALPix}}
\newcommand{\healpix}{\HEALPix}

\newcommand{\emcee}{\code{emcee}}
\newcommand{\ugali}{\code{ugali}}
\newcommand{\simple}{\code{simple}}
\newcommand{\var}[1]{\ensuremath{\texttt{\MakeUppercase{#1}}}\xspace}

\providecommand\physrep{\ref@jnl{Phys.~Rep.}}%
\providecommand\apjs{\ref@jnl{ApJS}}%
\providecommand{\jcap}{\ref@jnl{JCAP}}%

\shorttitle{Six More Ultra-Faint Milky Way Stellar Systems Discovered by DELVE}
\shortauthors{DELVE Collaboration}

\begin{document}

\reportnum{\footnotesize FERMILAB-PUB-22-704-LDRD-PPD}

\title{Six More Ultra-Faint Milky Way Companions Discovered in the DECam Local Volume Exploration Survey}


\author[0000-0003-1697-7062]{W.~Cerny}
\affiliation{Kavli Institute for Cosmological Physics, University of Chicago, Chicago, IL 60637, USA}
\affiliation{Department of Astronomy and Astrophysics, University of Chicago, Chicago IL 60637, USA}
\affiliation{Department of Astronomy, Yale University, New Haven, CT 06520, USA}

\author[0000-0002-9144-7726]{C.~E.~Mart\'inez-V\'azquez}
\affiliation{Gemini Observatory, NSF's NOIRLab, 670 N. A'ohoku Place, Hilo, HI 96720, USA}

\author[0000-0001-8251-933X]{A.~Drlica-Wagner}
\affiliation{Fermi National Accelerator Laboratory, P.O.\ Box 500, Batavia, IL 60510, USA}
\affiliation{Kavli Institute for Cosmological Physics, University of Chicago, Chicago, IL 60637, USA}
\affiliation{Department of Astronomy and Astrophysics, University of Chicago, Chicago IL 60637, USA}

\author[0000-0002-6021-8760]{A.~B.~Pace}
\affiliation{McWilliams Center for Cosmology, Carnegie Mellon University, 5000 Forbes Ave, Pittsburgh, PA 15213, USA}

\author[0000-0001-9649-4815]{B.~Mutlu-Pakdil}
\affiliation{Department of Physics and Astronomy, Dartmouth College, Hanover, NH 03755, USA}

\author[0000-0002-9110-6163]{T.~S.~Li}
\affiliation{Department of Astronomy and Astrophysics, University of Toronto, 50 St. George Street, Toronto ON, M5S 3H4, Canada}

\author[0000-0001-5805-5766]{A.~H.~Riley}
\affiliation{Institute for Computational Cosmology, Department of Physics, Durham University, South Road, Durham DH1 3LE, UK}
\affiliation{George P. and Cynthia Woods Mitchell Institute for Fundamental Physics and Astronomy,\\ Texas A\&M University, College Station, TX 77843, USA}
\affiliation{Department of Physics and Astronomy, Texas A\&M University, College Station, TX 77843, USA}

\author[0000-0002-1763-4128]{D.~Crnojevi\'c}
\affiliation{Department of Chemistry and Physics, University of Tampa, 401 West Kennedy Boulevard, Tampa, FL 33606, USA}
\author[0000-0003-4383-2969]{C.~R.~Bom}
\affiliation{Centro Brasileiro de Pesquisas F\'isicas, Rua Dr. Xavier Sigaud 150, 22290-180 Rio de Janeiro, RJ, Brazil}
\author[0000-0002-3690-105X]{J.~A.~Carballo-Bello}
\affiliation{Instituto de Alta Investigaci\'on, Sede Esmeralda, Universidad de Tarapac\'a, Av. Luis Emilio Recabarren 2477, Iquique, Chile}
\author[0000-0002-3936-9628]{J.~L.~Carlin}
\affiliation{Rubin Observatory/AURA, 950 North Cherry Avenue, Tucson, AZ, 85719, USA}
\author[0000-0002-7155-679X]{A.~Chiti}
\affiliation{Department of Astronomy and Astrophysics, University of Chicago, Chicago IL 60637, USA}
\affiliation{Kavli Institute for Cosmological Physics, University of Chicago, Chicago, IL 60637, USA}
\author[0000-0003-1680-1884]{Y.~Choi}
\affiliation{Department of Astronomy, University of California, Berkeley, Berkeley, CA 94720, USA}
\author[0000-0002-1693-3265]{M.~L.~M.~Collins}
\affiliation{Department of Physics, University of Surrey, Guildford GU2 7XH, UK}
\author[0000-0002-8800-5652]{E~Darragh-Ford}
\affiliation{Department of Physics, Stanford University, 382 Via Pueblo Mall, Stanford, CA 94305, USA}
\affiliation{Kavli Institute for Particle Astrophysics \& Cosmology, P.O.\ Box 2450, Stanford University, Stanford, CA 94305, USA}
\affiliation{SLAC National Accelerator Laboratory, Menlo Park, CA 94025, USA}
\author[0000-0001-6957-1627]{P.~S.~Ferguson}
\affiliation{Department of Physics, University of Wisconsin-Madison, Madison, WI 53706, USA}
\author[0000-0002-7007-9725]{M.~Geha}
\affiliation{Department of Astronomy, Yale University, New Haven, CT 06520, USA}
\author{D.~Mart\'{i}nez-Delgado}
\affiliation{Instituto de Astrof\'{i}sica de Andaluc\'{i}a, CSIC, E-18080 Granada, Spain}
\author[0000-0002-8093-7471]{P.~Massana}
\affiliation{Department of Physics, Montana State University, P.O. Box 173840, Bozeman, MT 59717-3840}
\author[0000-0003-3519-4004]{S.~Mau}
\affiliation{Department of Physics, Stanford University, 382 Via Pueblo Mall, Stanford, CA 94305, USA}
\affiliation{Kavli Institute for Particle Astrophysics \& Cosmology, P.O.\ Box 2450, Stanford University, Stanford, CA 94305, USA}
\author[0000-0003-0105-9576]{G.~E.~Medina}
\affiliation{Astronomisches Rechen-Institut, Zentrum f{\"u}r Astronomie der Universit{\"a}t Heidelberg, M{\"o}nchhofstr. 12-14, 69120 Heidelberg, Germany}
\author[0000-0002-0810-5558]{R.~R.~Mu\~{n}oz}
\affiliation{Departamento de Astronom\'ia, Universidad de Chile, Camino El Observatorio 1515, Las Condes, Santiago, Chile}
\author[0000-0002-1182-3825]{E.~O.~Nadler}
\affiliation{Observatories of the Carnegie Institution for Science, 813 Santa Barbara St., Pasadena, CA 91101, USA}
\affiliation{Department of Physics \& Astronomy, University of Southern California, Los Angeles, CA, 90007, USA}
\author[0000-0002-7134-8296]{K.~A.~G.~Olsen}
\affiliation{NSF's National Optical-Infrared Astronomy Research Laboratory, 950 N. Cherry Ave., Tucson, AZ 85719, USA}
\author[0000-0001-9186-6042]{A.~Pieres}
\affiliation{Laborat\'orio Interinstitucional de e-Astronomia - LIneA, Rua Gal. Jos\'e Cristino 77, Rio de Janeiro, RJ - 20921-400, Brazil}
\affiliation{Observat\'orio Nacional, Rua Gal. Jos\'e Cristino 77, Rio de Janeiro, RJ - 20921-400, Brazil}
\author[0000-0002-1594-1466]{J.~D.~Sakowska}
\affiliation{Department of Physics, University of Surrey, Guildford GU2 7XH, UK}
\author[0000-0002-4733-4994]{J.~D.~Simon}
\affiliation{Observatories of the Carnegie Institution for Science, 813 Santa Barbara St., Pasadena, CA 91101, USA}
\author[0000-0003-1479-3059]{G.~S.~Stringfellow}
\affiliation{Center for Astrophysics and Space Astronomy, University of Colorado, 389 UCB, Boulder, CO 80309-0389, USA}
\author[0000-0003-4341-6172]{A.~K.~Vivas}
\affiliation{Cerro Tololo Inter-American Observatory, NSF's National Optical-Infrared Astronomy Research Laboratory,\\ Casilla 603, La Serena, Chile}
\author{A.~R.~Walker}
\affiliation{Cerro Tololo Inter-American Observatory, NSF's National Optical-Infrared Astronomy Research Laboratory,\\ Casilla 603, La Serena, Chile}
\author[0000-0003-2229-011X]{R.~H.~Wechsler}
\affiliation{Department of Physics, Stanford University, 382 Via Pueblo Mall, Stanford, CA 94305, USA}
\affiliation{Kavli Institute for Particle Astrophysics \& Cosmology, P.O.\ Box 2450, Stanford University, Stanford, CA 94305, USA}
\affiliation{SLAC National Accelerator Laboratory, Menlo Park, CA 94025, USA}

\collaboration{32}{(DELVE Collaboration)}

\correspondingauthor{William Cerny}
\email{william.cerny@yale.edu}

\vspace{2.0em}
\begin{abstract}
We report the discovery of six ultra-faint Milky Way satellites discovered through matched-filter searches conducted using Dark Energy Camera (DECam) data processed as part of the second data release of the DECam Local Volume Exploration (DELVE) survey. Leveraging deep Gemini/GMOS-N imaging (for four candidates) as well as follow-up DECam imaging (for two candidates), we characterize the morphologies and stellar populations of these systems.   We find that these candidates all share faint absolute magnitudes ($M_V \geq -3.2$\,mag) and old, metal-poor stellar populations ($\tau > 10$\,Gyr,  $\rm [Fe/H] < -1.4$\,dex).  Three of these systems are more extended ($r_{1/2} > 15$\,pc), while the other three are compact ($r_{1/2} < 10$\,pc). From these properties, we infer that the former three systems (Bo\"{o}tes~V, Leo Minor I, and Virgo II) are consistent with ultra-faint dwarf galaxy classifications, whereas the latter three (DELVE 3, DELVE 4, and DELVE 5) are likely ultra-faint star clusters. Using data from the \Gaia satellite, we confidently measure the proper motion of Bo\"{o}tes~V, Leo Minor I, and DELVE~4, and tentatively detect a proper motion signal from DELVE~3 and DELVE~5; no signal is detected for Virgo II. We use these measurements to explore possible associations between the newly-discovered systems and the Sagittarius dwarf spheroidal, the Magellanic Clouds, and the Vast Polar Structure, finding several plausible associations.
Our results offer a preview of the numerous ultra-faint stellar systems that will soon be discovered by the Vera C.\ Rubin Observatory and highlight the challenges of classifying the faintest stellar systems.

\end{abstract}


\section{Introduction}
\label{sec:intro}
The standard picture of galaxy formation within the Lambda-Cold-Dark-Matter ($\Lambda$CDM) paradigm describes that galaxies form from the condensation of gas onto dark matter ``halos,'' which grow over cosmic time through merging and accretion of lower-mass ``subhalos'' \citep[e.g.,][]{1978MNRAS.183..341W, 1984Natur.311..517B, 2001MNRAS.321..372J, 2008MNRAS.391.1685S, 2010AdAst2010E...8K}. As a result of this hierarchical formation channel, the dark matter halos hosting massive galaxies like the Milky Way are expected to be rich with substructure, boasting numerous subhalos, which can host their own small galaxies \citep[e.g.,][]{2007ApJ...667..859D, 2008MNRAS.391.1685S, 2010AdAst2010E...8K,2016ApJ...818...10G}. The least massive and least luminous galaxies in this hierarchy are known as ``ultra-faint'' dwarf galaxies (see \citealt{2019ARA&A..57..375S} for a recent review). These extreme systems are believed to have originated within some of the earliest halos to collapse in the universe, and likely formed multiple generations of stars quickly before being quenched during or before the epoch of reionization \citep{2014ApJ...796...91B,2021ApJ...920L..19S}. 
The dark matter halos that host these galaxies provided sufficiently deep gravitational potential wells to retain ejecta from supernovae, and, by virtue of their resident galaxies' lack of extended star formation, preserve a fossil record of their chemical enrichment histories \citep{2009ApJ...693.1859B,2012AJ....144...76W,2012ApJ...759..115F}. This formation and evolutionary pathway makes these galaxies nearly pristine laboratories for studying the nature of dark matter, tracing the births and deaths of the first stars, and studying the physics of the early universe.

\par Although ultra-faint dwarf galaxies are expected to be the most numerous class of galaxy in the universe, their diffuse morphologies and exceedingly low luminosities rendered them undetectable until relatively recently. It was not until the advent of the Sloan Digital Sky Survey (SDSS; \citealt{2000AJ....120.1579Y}) and the application of automated matched-filter search algorithms that the first exemplar of these systems was detected \citep{2005AJ....129.2692W}. Since these early efforts in SDSS, numerous deep, wide-area surveys including the Panoramic Survey Telescope and Rapid Response System 1 survey (PS1; \citealt{2016arXiv161205560C}), the VLT Survey Telescope ATLAS survey \citep{2015MNRAS.451.4238S}, the Dark Energy Survey (DES; \citealt{2005astro.ph.10346T}), the Hyper Suprime-Cam Subaru Strategic Program survey (HSC SSP; \citealt{hscssp}), and the DECam Local Volume Exploration survey (DELVE; \citealt{2021ApJS..256....2D}) have enabled the rapid discovery of ultra-faint galaxies around the Milky Way. Recently, the European Space Agency \Gaia satellite has also offered the ability to detect co-moving structures in the stellar halo, expanding the detectable parameter space of dwarf galaxies to larger physical sizes and lower surface brightness thresholds \citep{2019MNRAS.488.2743T, 2021ApJ...915...48D}. The census of ultra-faint dwarf galaxies has now advanced to more than sixty systems, nearly all of which have been identified as resolved satellites of the Milky Way, M31/Andromeda, and a select few nearby hosts in the Local Volume \citep[][and references therein]{McConnachie:2012, 2019ARA&A..57..375S}. These systems continue to be discovered at a rapid pace, with roughly ten ultra-faint dwarf galaxy candidates discovered in the last few years alone \citep[e.g., see][]{2019PASJ...71...94H,2020ApJ...890..136M, 2021ApJ...920L..44C, 2022arXiv220311788C,2022MNRAS.509...16M,2022ApJ...926...77M, 2022MNRAS.515L..72C,2022ApJ...935L..17S,2022ApJ...937L...3B}.
\par In tandem with recent dwarf galaxy discoveries, systematic searches for ultra-faint systems have also uncovered an enigmatic population of faint, compact, and ancient stellar systems with $M_V \gtrsim -3$, azimuthally-averaged half-light radii $r_{1/2} < 15$ pc, and heliocentric distances $D_{\odot} \gtr 10$ kpc \citep[e.g.,][]{2011AJ....142...88F,2012ApJ...753L..15M,2013ApJ...767..101B,2017MNRAS.470.2702K, 2019MNRAS.484.2181T, 2020ApJ...890..136M}.  These systems have typically been classified as ultra-faint halo star clusters on the basis of their lower luminosities, smaller sizes, and more spherical stellar density distributions relative to the known population of ultra-faint dwarf galaxies. Under this interpretation, these systems may have originated within low-mass galaxies that disrupted while accreting onto the Milky Way, as is believed to be the case for at least some of the Milky Way's ``classical'' globular clusters \citep{1978ApJ...225..357S}. Although these systems likely do not exhibit the high dark matter densities and mass-to-light ratios of their dwarf galaxy counterparts, many of these clusters may have nonetheless survived the strong tidal forces exerted on them by the Milky Way by virtue of their compactness. This all being said, the relatively small number of known ultra-faint halo star clusters, combined with the near-complete lack of spectroscopic radial velocity and velocity dispersion measurements for these systems, has limited our ability to assess both their dark matter contents and their origins.
\par Here, we present the discovery of six ultra-faint Milky Way satellite systems spanning an intermediate range of physical sizes encompassing both the star cluster and dwarf galaxy regimes. In \secref{data_and_search}, we present an overview of the DELVE DR2 dataset and our satellite search procedure. In \secref{followup}, we describe our follow-up imaging of these candidates using the Gemini Multi-Object Spectrograph at Gemini North (GMOS-N) and the Dark Energy Camera (DECam). We use this imaging data to characterize the morphologies and stellar populations of these systems in \secref{properties}, as well as measure their proper motions using data from the \Gaia satellite. In \secref{review} and \secref{association}, we discuss the nature and classifications of these individual systems and their potential associations with various Local Group structures. We conclude in \secref{summary}.
\section{DELVE Data and Satellite Search}
\label{sec:data_and_search}
\subsection{DELVE DR2}
We conducted a search for arcminute-scale stellar substructures in the Milky Way halo using data from the second data release of DELVE (DELVE DR2; \citealt{delvedr2}).\footnote{https://datalab.noirlab.edu/delve/} DELVE seeks to achieve contiguous, multi-band coverage of the entire high-Galactic-latitude sky accessible to DECam (\citealt{2010SPIE.7735E..0DF}) by combining 126 nights of new DECam observations in the $g,r,i,z$ bands with existing archival data from numerous NOIRLab community programs.  DELVE processes all exposures with the Dark Energy Survey Data Management pipeline (DESDM; \citealt{2018PASP..130g4501M}), and derives astrometric and photometric measurements from individual exposures, as opposed to making measurements on stacked images. The catalogs from the individual exposures are matched across filters and merged to form a single unified multi-band catalog following the procedure described in \citet{2021ApJS..256....2D}.
\par
\begin{figure*}
    \centering
    \includegraphics[width = .8\textwidth]{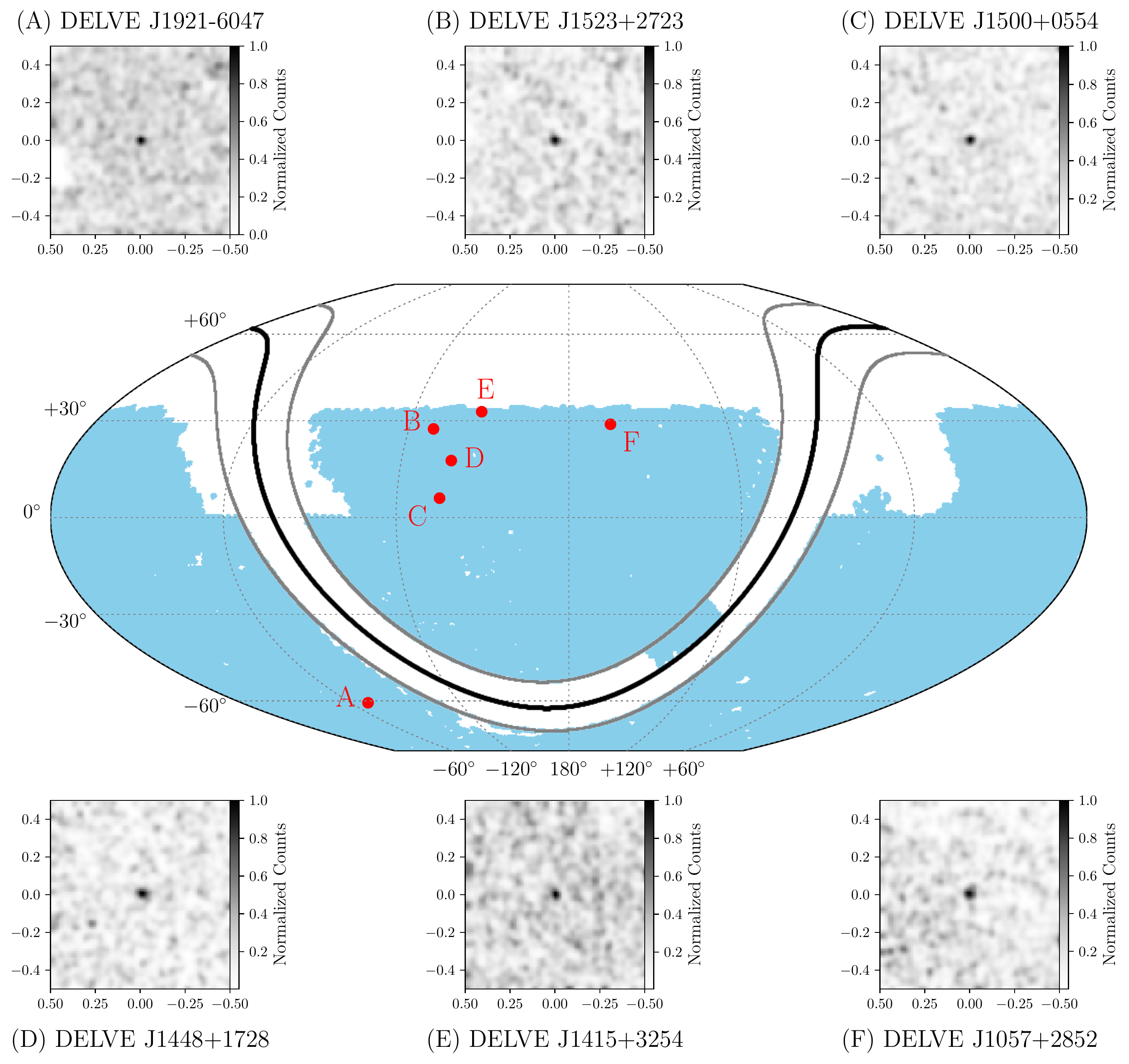}
    \caption{Skymap of the DELVE DR2 $g$-band coverage (blue background), shown in McBryde-Thomas flat-polar quartic projection. The Galactic plane ($b = 0\degree$) is shown as a solid black curve, and two curves denoting $b = \pm 10\degree$ are shown in grey. The locations of the candidates reported in this work are shown as red dots, labelled by decreasing right ascension. Each inset panel shows the isochrone-filtered stellar density field for a 1.0$\degree \times 1.0\degree$ region centered on a given candidate, smoothed with a 1$\arcmin$ Gaussian kernel. These panels were generated exclusively using DELVE DR2 data, and do not include the follow-up exposures from DECam and GMOS-N described in \secref{followup}. These panels include only sources with $g_0, r_0 < 24$ mag and with $0 \leq \var{EXTENDED\_CLASS\_G} \leq 2$ (see \citealt{delvedr2} for details on this morphological classifier).}
    \label{fig:mapfigure}
\end{figure*}

 The most recent iteration of the DELVE photometric catalog,  DELVE DR2, was derived from $\sim160,000 $ exposures spanning a sky area of $20,000 - 25,000 \text{ deg}^2$ in each band, with $>17,000 \text{ deg}^2$ of simultaneous four-band coverage. The DECam programs contributing the largest number of exposures to DELVE DR2 include DELVE itself (Proposal ID: 2019A-0305), DES \citep{2021ApJS..255...20A}, the DECam Legacy Surveys (DECaLS; \citealt{Dey:2019}), and the DECam eROSITA Survey (DeROSITAS; Proposal ID: 2021A-0149).
As depicted for the $g$-band in \figref{mapfigure}, the full DELVE DR2 dataset spans nearly the entire southern sky with Galactic latitude $|b| > 10\degree$ and declination $\delta_{2000} <0\degree$, as well as $\sim 2/3$ of the sky above the celestial equator with $|b| > 10\degree$ and $\delta_{2000} \lesssim 33\degree$. The median $10\sigma$ point-source depths for DELVE DR2 are $g=23.5$, $r=23.1$, $i=22.7$ and $z=22.1$ mag, although these vary as a function of sky position depending on the available archival data.
We refer the reader to \citet{2021ApJS..256....2D} and \citet{delvedr2} for a more complete description of the DELVE science goals, observing strategy, data processing pipeline, and data products.

\par We corrected the DELVE DR2 photometry, as well as all other photometry presented in this work, for reddening due to interstellar dust by interpolating the $E(B-V)$ values from the maps of \citet{1998ApJ...500..525S} with the rescaling from \citet{2011ApJ...737..103S}. We then calculated bandpass-specific extinctions using the total-to-selective absorption coefficients from DES DR1 \citep{2018ApJS..239...18A}. Throughout this work,  all reported magnitudes refer to AB magnitudes in the DECam photometric system unless otherwise noted, with extinction-corrected magnitudes denoted by a subscript ``0''.

\subsection{Satellite Search}
\par Our satellite search consisted of running four iterations of the \code{simple} algorithm, which has previously been applied to data from multiple DECam survey programs, as well as data from PS1, to conduct an extensive census of Milky Way satellites \citep{2015ApJ...807...50B,2015ApJ...813..109D,2020ApJ...893...47D}.\footnote{\url{https://github.com/DarkEnergySurvey/simple}} The \code{simple} algorithm applies an isochrone matched-filter in  color-magnitude space defined by two filter bands to enhance the density contrast of halo substructures across a range of distances. 
\par We began by dividing the DELVE DR2 footprint into HEALPix \citep{Gorski:2005} pixels at $\var{NSIDE} = 32$, corresponding to an area of 3.4\,deg$^2$. We then selected all stars with colors consistent with a synthetic \citet{Bressan:2012} isochrone, which was scanned across a grid of distance moduli spanning $(m-M)_0 = 18.0$ to $(m-M)_0 = 21.5$  in steps of 0.25 mag (corresponding to $\sim 40$ kpc to $200$ kpc). Next, the resulting stellar density field was filtered with a Gaussian kernel ($\sigma = 2\arcmin$), and local overdensities were identified by iteratively raising a density threshold until fewer than ten disconnected peaks remained. Each overdensity was assigned a significance based on the Poisson significance of the observed stellar counts compared to the field density estimated from a concentric background annulus.
\par To maximize the discovery potential of the search, we specifically ran four iterations of the procedure above. Two of these iterations ran over all regions of DELVE DR2 with simultaneous $g,r$ band coverage, while the other two covered regions with $g,i$ band data. For each filter pair, we performed a single search using an old, metal-poor PARSEC isochrone \citep{Bressan:2012} with $\tau = 12 \Gyr;$  $Z = 0.0001$ ($\rm [Fe/H] \sim -2.2$\,dex) and another with a slightly more metal-rich isochrone with $\tau = 12 \Gyr$; $Z = 0.0005$ ($\rm [Fe/H] \sim -1.5$\,dex). These searches individually produced thousands of overdensities above our nominal significance threshold of $5.5\sigma$. We proceeded to generate diagnostic plots for this composite sample of overdensities, from which we identified 54 that merited further investigation on the basis of their smoothed stellar density distributions (as in \figref{mapfigure}), color--magnitude diagrams, or appearance in the color images available through the Legacy Surveys Sky Viewer (see \secref{review} for more details).  Many of these 54 overdensities were independently detected across two or more iterations of our search (e.g., in the $g,r$ search and the $g,i$ search), suggesting that the signals at these locations were robust to at least some of our search systematics.\footnote{We did not require overdensities to be detected in multiple iterations of our search because the sky regions covered by the $r$ and $i$ bands differ considerably. However, the final six candidates that we report were each detected in $\geq 2$ of the search iterations (see \secref{review}).}
\par We ran a full Markov Chain Monte Carlo (MCMC) parameter fit to the structure and stellar population of all 54 overdensities using the Ultra-faint GALaxy LIkelihood software toolkit (\ugali; see \secref{decam_properties}). For a significant majority of these systems, the MCMC failed to converge or converged to a clearly non-physical outcome.\footnote{Most often, the failure involved the MCMC converging to the maximum size allowed by our prior bounds. Because the small kernel size used in the search was specifically chosen to detect compact (arcminute-scale) overdensities, these cases could safely be rejected as false positive detections.} This allowed us to further narrow the sample to seven candidates which we believed to be genuine stellar associations in the Milky Way halo (i.e., dwarf galaxies or star clusters). We were able to acquire deeper follow-up data for six of these systems, which form the basis of this work. We were unable to constrain the properties of the seventh candidate system given the shallow depth of the available data, and we intend to revisit it at a later date when follow-up observations become available.

\par \figref{mapfigure} shows the spatial distribution of the candidates reported in this work (red dots), overlaid on the DELVE DR2 $g$-band footprint (light blue). Although we do not seek to quantify our search sensitivity and completeness in this paper, we do observe that five of our six candidates lie in regions of sky at $\delta_{2000} > 0\degree$ covered by DECaLS. There are at least two possible interpretations of the spatial distribution of our candidates. Firstly, DELVE has already reported four satellite systems south of the equator and one in the north at $\alpha_{2000} \sim 330\degree$; by consequence of these previous searches, subsequent discoveries in DELVE are less likely to fall in those regions. Secondly, the Legacy Surveys Sky Viewer made it possible to quickly inspect color images of each candidate in these regions, which proved especially useful for assessing the presence of compact stellar systems; a similar tool is not yet available for DECam data in the south outside of DES. The final candidate falls just outside of the DES footprint, suggesting that additional candidates may yet be discovered even in previously-inspected regions of the DELVE survey.

\begin{deluxetable*}{l c c c c c c c c}
\tablecolumns{6}
\tablewidth{0pt}
\tabletypesize{\small}
\tablecaption{\label{tab:obsTable} Summary of Follow-up Observations Presented in this Work}
\tablehead{\colhead{Candidate} & \colhead{Instrument} & \colhead{Exposure Times} & \colhead{Dates}  & \colhead{Adopted Magnitude Limit}}
\startdata
DELVE J1921-6047 (DELVE~3)& DECam &  ($g$) 3 x 300s ($r$) 3 x 300s & 2022-07-28 &$g_0 = 23.5$; $r_0 = 23.1$\\
DELVE J1523+2723 (DELVE~4) & GMOS-N & ($g$) 5 x 300s ; ($r$) 4 x 300s  & \text{2022-07-04}   &   $g_0 = 25.09$; $r_{0} = 24.69$ \\
DELVE J1500+0554 (Virgo~II)& GMOS-N & ($g$) 12 x 150s; ($r$) 10 x 150s  & \text{2022-07-20}   &  $g_{0} = 26.30; r_{0} = 25.79$ &   \\
DELVE J1448+1728 (DELVE~5) & GMOS-N & ($g$) 10 x 300s ; ($r$) 3 x 300s  &    2022-07-31 & $g_{0} = 25.20; r_{0} = 24.69$ \\
DELVE J1415+3254 (Bo\"{o}tes~V) & GMOS-N &  ($g$) 8 x 300s ; ($r$) 5 x 350s  & 2022-06-27  &  $g_{0} = 25.87; \ r_{0} = 25.51$ \\
DELVE J1057+2852 (Leo~Minor~I) & DECam & ($g$) 3 x 300s; ($r$) 3 x 300s & 2022-05-21 & $g_0 = 23.4$; $r_0 = 23.1$\\
\enddata
\tablecomments{The DECam exposures reported here represent only the follow-up exposures that we collected; however, these were combined with existing archival observations to generate the catalog we used throughout this work. For the candidates with GMOS-N data, this table reports only the exposures that ultimately became part of our photometric catalogs, and therefore excludes some exposures with lower image quality that were taken as part of our program (especially $g$-band exposures targeting DELVE~4). We note that additional $r$-band observations of DELVE~5 were planned but not executed due to Gemini queue constraints. The DECam magnitude limits correspond to the $10\sigma$ depth and the GMOS-N magnitude limits corresponding to the 30\% completeness level based on the artificial star tests described in \secref{gmosphotometry}.}
\end{deluxetable*}
During the preparation of this manuscript, the authors of \citet{Smith:2022} informed us of their independent discovery of the stellar system referred to here as Bo\"otes~V using data from the Ultraviolet Near-Infrared Optical Northern Survey (UNIONS).\footnote{\url{https://www.skysurvey.cc/}} 
We have not altered our analysis of this system in light of their work, nor have we reviewed their manuscript. 
We also note that this system, as well as the system that we name DELVE~4, is included in the candidate lists from \citet{2021ApJ...915...48D}, who performed a search for faint Milky Way satellites by applying a wavelet transform approach to \Gaia proper motion data; however, the authors did not specifically isolate these systems for study. 
We encourage the reader to review these works for details about these complementary search efforts.

\section{Follow-up Imaging of the Newly-Discovered Candidates with DECam and Gemini/GMOS-N}
\label{sec:followup}

We pursued deeper follow-up imaging of each newly-discovered candidate with either DECam on the 4-m Blanco Telescope or GMOS-N (\citealt{1997SPIE.2871.1099D,2004PASP..116..425H}) on the 8.1-m Gemini Telescope during late spring and summer of 2022. The goal of these observations was to increase the number of putative member stars in each system in order to confirm whether they represented \textit{bona fide} associations of stars, and, if so, to improve constraints on their stellar populations and structural properties.

\par In brief, this follow-up imaging consisted of (1) a 4-hour Gemini Fast Turnaround program targeting DELVE J1523+2723, DELVE J1500+0554, DELVE J1148+1728, (2) a 1.4-hour Gemini Director's Discretionary program studying DELVE J1415+3254, (3) $30$ minutes of DECam imaging targeting DELVE J1057+2852 during regular DELVE observing time, and (4) $30$ minutes of DECam imaging targeting DELVE 1921-6047, again during regular DELVE time. A summary of these observations is presented in \tabref{obsTable}. We describe the corresponding data reduction procedures in the following sections.

\subsection{GMOS-N Data Reduction and Photometry}
\label{sec:gmosphotometry}

\par We reduced the GMOS-N data using a lightly-modified version of the standard GMOS-N imaging reduction pipeline implemented by the Data Reduction for Astronomy from Gemini Observatory North and South platform (DRAGONS; \citealt{2019ASPC..523..321L}).\footnote{\url{https://dragons.readthedocs.io/en/stable/}} The DRAGONS recipe for GMOS-N image reduction automatically performs overscan and bias subtraction, flat-fielding, image stacking, and cosmic ray rejection based on the stacked image. Because Gemini operates on a queue, we did not collect our own flat and bias frames; instead, we used the most recently available flat and bias frames in the Gemini archive.

\par We performed point-spread-function (PSF) photometry on the resultant $g$ and $r$ stacked images using the {\sc daophot iv/allstar} suite of codes \citep{1987PASP...99..191S}. An empirical PSF was derived for each image using $\sim$50 stars that were not saturated, far from the edges of the CCDs, and spread over the whole field-of-view. A Moffat function ($\beta=2.5$) was the preferred PSF model over the whole set of images since it provided the smallest residuals. Following the same procedure as in \citet{Martinez-Vazquez:2021a}, we performed initial photometry on the stacked images with a spatially-constant PSF using {\sc allstar}; this provided both a preliminary photometric catalog and a star-subtracted image. Because some of the faintest sources were undetected in this first pass, we identified these faint sources using the star-subtracted image and appended them to our previous catalog in order to obtain a more complete catalog for every image. This strategy enables the recovery of faint sources located in the PSF wings of brighter objects \citep[e.g.,][]{Martinez-Vazquez:2021a, Cantu:2021}. Finally, after subtracting the PSF stars from the stacked images, we refined the PSF model and also allowed a quadratic variation of the PSF throughout the field-of-view (FOV) in order to perform the final PSF photometry of our catalog. 
\par We then calibrated the instrumental GMOS-N photometry into the DECam photometric system using the available data from DELVE DR2, including a color term to account for the difference between the DECam and GMOS-N filter systems. We explored a range of calibration techniques, and found that deriving the color term from the photometry for a given candidate provided a better relative calibration than a simultaneous fit to the photometry for all four candidates with GMOS-N data; however, no solution was found to be perfect. To account for lingering differences between the DELVE DR2 photometry and the calibrated GMOS-N photometry, we assume a 0.03 mag photometric error floor for all GMOS-N magnitudes quoted in this work. This primarily affects a small number of brighter stars in each candidate, as the photometric errors for fainter sources are dominated by photon shot noise (as quantified via the artificial star tests described below).
\par To determine the completeness and photometric errors of our GMOS-N data as a function of magnitude, we performed artificial star tests following a similar approach to that described in \citet{Monelli:2010}. Approximately 30,000 artificial stars were injected into each stacked GMOS-N image spread over a regular grid covering the FOV.  Artificial stars were placed at the vertices of equilateral triangles separated by a distance of twice the PSF radius plus one pixel. This strategy allowed us to add the artificial stars in a way that avoids overcrowding. Depending on the image quality of each stacked image, we injected between 1160 to 1900 artificial stars per iteration and per filter, for 16 to 25 iterations. The colors and magnitudes of the artificial stars were obtained using IAC-BaSTI synthetic color--magnitude diagrams \citep{Hidalgo:2018, Pietrinferni:2021}\footnote{\url{http://basti-iac.oa-abruzzo.inaf.it/syncmd.html}} and covered the full range of our observed color--magnitude diagrams.
\par We display the completeness as a function of magnitude in \appref{complete}, and we report the magnitude corresponding to the 30\% completeness limit in \tabref{obsTable}. These values were calculated after making the same photometric quality cuts used in our isochrone-fitting and structural property analyses (see \secref{properties}). We opted to extend our stellar selection to the 30\% completeness limit when fitting the distance and structure of the systems observed by GMOS-N to ensure sufficient contrast relative to the background for these faint systems. When deriving the absolute magnitude of each system, we applied a more elaborate completeness correction as a function of magnitude (see \secref{absmag}).

\subsection{Additional Exposures from DECam}

We collected additional 300s exposures for DELVE J1057+2852 and DELVE J1921-6047 with DECam in late May 2022 and August 2022, respectively. Because of the large declination of DELVE J1057+2852, we were only able to observe it at airmass $\roughly 1.9$. In conjunction with the poor sky conditions, this conspired to make the new observations comparable in quality to the existing data in DELVE DR2. These observations did, however, fill in the CCD gaps in the $g$-band coverage from DELVE DR2 (see \figref{mapfigure}). Although we were able to observe DELVE J1921-6047 at low airmass, the variable sky conditions likewise limited our data quality in $g$-band. By contrast, a brief window of clear sky allowed us to improve upon the existing $r$-band data for this system.
\par We combined these follow-up exposures with the exposures available at the time of discovery to construct new multi-band source catalogs following the identical procedure used to generate DELVE DR2.  For the analyses described hereafter, we opted to use PSF magnitudes derived from a weighted-average of measurements derived from individual exposures ($\var{WAVG\_MAG\_PSF\_G}$ and $\var{WAVG\_MAG\_PSF\_R}$; see Appendix A of \citealt{2021ApJS..255...20A} for details). This allowed us to leverage our follow-up exposures to achieve better photometric precision even in cases where these exposures did not improve upon the seeing or the effective exposure time of the discovery data. In practice, this contributed to slightly better depth for our final catalog compared to DELVE DR2.

\section{Properties of the Newly-Discovered Systems}
\label{sec:properties}
We constrained the morphological and stellar population properties of the six candidates through maximum-likelihood analyses using the GMOS-N data (for DELVE J1523+2723, DELVE J1500+0554, DELVE J1448+1728, and DELVE J1057+2852) and the deeper DECam data (for DELVE J1921-6047 and DELVE J1057+2852). Although previous DELVE analyses -- all of which have been based on DECam data -- have exclusively used the \ugali toolkit to characterize the structure and stellar populations of ultra-faint stellar system candidates, the smaller size of the GMOS-N FOV ($5.5\arcmin \times 5.5\arcmin$) demanded that we adopt a different approach. More specifically, the smaller GMOS-N FOV would make it difficult to define a suitably large background annulus to define a foreground model for the \ugali joint fit of color, magnitude, and spatial distributions. Therefore, we used \ugali only for the candidates with DECam data, and describe our alternate procedure for fitting the candidates with GMOS-N data below.

\begin{deluxetable*}{l c c c c c c c}
\tablecolumns{7}
\tablewidth{0pt}
\tabletypesize{\small}
\tablecaption{\label{tab:properties} Morphological, Stellar Population, and Proper Motion Properties of the Newly-Discovered Systems}
\tablehead{\colhead{Parameter} &  \colhead{Unit}  &  \colhead{DELVE 3}  &  \colhead{DELVE 4} &  \colhead{Virgo II} & \colhead{DELVE 5} & \colhead{Bo\"{o}tes~V} & \colhead{Leo Minor I}}
\startdata
IAU Name & ... &  J1921-6047 & J1523+2723 & J1500+0554 & J1448+1728 & J1415+3254 & J1057+2852 \\
Instrument & ... & DECam & GMOS & GMOS & GMOS &  GMOS & DECam \\
\ra   & deg  & $290.396^{+0.003}_{-0.003}$ & $230.775^{+0.001}_{-0.001}$ & $225.059^{+0.001}_{-0.001}$ & $222.104^{+0.002}_{-0.002}$&  $213.909^{+0.001}_{-0.001}$ & $164.261^{+0.005}_{-0.006}$\\
\dec  & deg  &  $-60.784^{+0.001}_{-0.001}$ & $27.395^{+0.001}_{-0.001}$ & $5.909^{+0.001}_{-0.001}$ & $17.468^{+0.001}_{-0.001}$ & $32.914^{+0.001}_{-0.001}$ & $28.875^{+0.004}_{-0.003}$ \\
$a_h$ & arcmin & -----  & $0.49^{+0.16}_{-0.12}$ & -----  & $0.68^{+0.24}_{-0.17}$& $0.76^{+0.08}_{-0.07}$  &  -----    \\
$a_{1/2}$ & pc   & -----  & $6^{+2}_{-2}$ & -----  & $8^{+3}_{-2} $ & $23^{+3}_{-3}$ &  -----  \\
$r_h$ & arcmin   &  $0.40^{+0.12}_{-0.08}$ & $0.36^{+0.09}_{-0.07}$ & $0.74^{+0.13}_{-0.11}$ & $0.45^{+0.12}_{-0.09}$ & $0.67^{+0.07}_{-0.06}$ & $1.09^{+0.37}_{-0.35}$\\
$r_{1/2}$ & pc   &  $7^{+2}_{-2} $ & $5^{+1}_{-1}$ &  $16^{+3}_{-3}$ & $5^{+1}_{-1}$ & $20^{+2}_{-2}$ &  ${26^{+9}_{-9}}$ \\
$\epsilon$\tablenotemark{\small a} & ... & $< 0.4 $ & $0.4^{+0.2}_{-0.2}$ & $< 0.3$ & $0.6^{+0.1}_{-0.2}$ & $0.2^{+0.1}_{-0.1}$  &  $< 0.4$  \\
$\PA$  & deg   & $87^{+30}_{-35}$ & $152^{+14}_{-17}$  & $131^{+44}_{-26}$ & $77^{+10}_{-11}$ & $18^{+13}_{-15}$ & $74^{+49}_{-39}$\\
$(m-M)_0$ & mag   & $18.73^{+0.09}_{-0.23}$ & $18.28^{+0.19}_{-0.19}$ & $19.30^{+0.22}_{-0.22}$ & $17.97^{+0.17}_{-0.17}$ & $20.04^{+0.15}_{-0.15} $& $19.56^{+0.11}_{-0.19}$\\
$D_{\odot}$ & kpc   & $56^{+2}_{-6}$ & $45^{+4}_{-4}$ & $72^{+8}_{-7}$ & $39^{+3}_{-3}$ & $102^{+7}_{-7}$ & $82^{+4}_{-7}$ \\
$\age$ & $\Gyr$ & $> 11.7$ & 13.5*& 11.0* & 10.0* & 12.5* & $> 11.4$ \\
${\rm [Fe/H]}$  & dex & $< -1.4$ & $-1.9$* & $-1.9$* & $-2.2$* & $-2.2$* & $< -1.9$ \\
$M_V$\tablenotemark{\small b} & mag   & $-1.3^{+0.4}_{-0.6}$ &  $-0.2^{+0.5}_{-0.8}$ & $-1.6^{+0.4}_{-0.6}$ & $0.4^{+0.4}_{-0.9}$ &  $-3.2^{+0.3}_{-0.3}$ & $-2.4^{+0.5}_{-0.4}$  \\
$\Sigma p_{i, \ugali}$\tablenotemark{\small c} &  ... & $25^{+5}_{-5}$ & ----- & ----- & ----- & ----- & $22 \pm 5$\\
$N_{\rm GMOS}$\tablenotemark{\small c} & ... & ----- & $33^{+8}_{-7}$ & $128^{+23}_{-20}$ & $33^{+8}_{-7}$ & $198^{+20}_{-19}$ & -----\\
$E(B-V)\tablenotemark{\small d}$ & mag & 0.054 & 0.042 & 0.039 & 0.030 & 0.013 & 0.018 \\ 
\hline
$\mu_{\alpha *}$ & mas yr$^{-1}$ & $-0.33_{-0.34}^{+0.31}$ & $+0.42_{-0.09}^{+0.08}$ & ... & $-1.82_{-0.12}^{+0.13}$ & $-0.22\pm0.05$ & $-0.01_{-0.40}^{+0.39}$ \\ 
$\mu_{\delta}$ & mas yr$^{-1}$ & $-0.80_{-0.32}^{+0.35}$ & $-0.75\pm0.11$ & ... & $-0.93\pm0.12$ & $-0.28\pm 0.07$ & $-1.29_{-0.40}^{+0.37}$  \\
$\sum_i p_{i, {\rm MM}}$ & ... & $2.0\pm 0.1$ & $6.0\pm 0.0$ & ... & $2.8\pm0.5$ & $6.1 \pm 0.1$ & $2.9\pm 0.1$ \\
\enddata
\tablecomments{Age and metallicity were not constrained in our GMOS-N analyses; instead, we fixed these parameters to the values reported in this table, and indicate these quantities with an asterisk. For DELVE 3 and Leo Minor I, the age and metallicity posteriors were distributed against the allowed bounds ($13.5 \Gyr$ and $\rm [Fe/H] \sim  -2.2$); we therefore quote upper and lower limits on the age and metallicity ($\tau$ and $\rm [Fe/H]$) at 84\% confidence.} 
\tablenotetext{a}{The posterior distribution for the ellipticity of DELVE 3, Virgo II, and Leo Minor I peaked near $\epsilon = 0$, and we therefore quote an upper limit at 84\% confidence. For these candidates, we therefore do not quote a value of the elliptical half-light radius ($a_h$ and $a_{1/2}$), and only quote the azimuthally-averaged half-light radius ($r_h$ and $r_{1/2}$).}
\tablenotetext{b}{The uncertainty in the absolute magnitude, $M_V$, was calculated following \citet{2008ApJ...684.1075M} and does not include the uncertainty on the distance.}
\tablenotetext{c}{$\Sigma p_{i, \ugali}$ and $N_{\rm GMOS}$ refer to the modeled number of stars above the magnitude limit for our DECam and GMOS-N analyses, respectively.}
\tablenotetext{d}{The $E(B-V)$ values listed here are calculated from the mean value of all DELVE DR2 sources within a three arcminute radius around the centroid of each candidate.}
\end{deluxetable*}

\subsection{Gemini/GMOS-N Analysis}

\par To fit the structural properties of the candidates with GMOS-N imaging, we applied a lightly modified version of the dwarf galaxy structural fitting code presented by \citet{2021ApJ...908...18S},\footnote{\url{https://github.com/jsimonastro/EriII-structural-fitting/}} which is a binned Poisson maximum-likelihood procedure based on the formalism presented in Appendix C of \citet{2020ApJ...893...47D}. Broadly, the parameter estimation procedure involved three main steps for each candidate: (1) creating a binned coverage mask defining the imaged area, (2) selecting a sample of probable member stars based on a simplified isochrone-fitting procedure, and (3) performing MCMC sampling to describe the posterior probability distribution of the structural parameters assuming a binned Poisson likelihood function. We elaborate upon each of these steps in the subsections below.
\subsubsection{Coverage Mask Definition}
\par For each candidate, we began by defining a full-resolution coverage map based on the stacked GMOS-N $g$-band image produced by DRAGONS. Image pixels within the imaging region of the frame were assigned a value of one, and regions of the focal plane not used for imaging were assigned zero; we found these regions were naturally separated by a threshold pixel value between 500 and 1400 electron counts, depending on the image. Next, we visually identified bright stars and the largest extended galaxies in the image and used a Python implementation of SourceExtractor, \code{SEP} \citep{2016JOSS....1...58B}, to define isophotal shape and orientation parameters, which we used to mask these objects.  We then binned this full-resolution mask into $25 \times 25$ pixel bins, assigning a coverage fraction between 0 and 1 representing the number of pixels in each bin that contained a value of one in the full-resolution mask. This represents a divergence from the procedure of \citet{2021ApJ...908...18S}, who used a binary (0 or 1) coverage map for their higher-resolution \textit{Hubble Space Telescope} data.
\subsubsection{Distance}
\par We performed a simple isochrone fitting procedure similar to that of \citet{2018ApJ...863...25M} to measure the distance to each candidate. We began by selecting a stellar sample by applying a cuts on the DAOPHOT indices $\chi$ and $sharp$, namely $\chi < 3$ and $-0.5 \leq sharp \leq 0.7$. We then applied a spatial selection for a given system based on an initial estimate of its angular half-light radius, specifically taking all stars within a circular selection of $r < 2r_h$.  Next, we adjusted the age and metallicity of a PARSEC isochrone to the resulting color--magnitude diagram, adopting an initial estimate of the distance provided by a $\ugali$ fit to the DECam data. 
We found that it was necessary to adjust the age and metallicity by eye due to the imperfect relative calibration, the dearth of red giant branch stars, and the generally large photometric errors at the main sequence turnoff for each system. 
We then iterated the isochrone with our chosen age and metallicity through a range of distance moduli spanning $\pm0.5$ mag about the initial guess in intervals of 0.01 mag.
At each step, we counted the number of observed stars in a given candidate with colors consistent with the isochrone according to $\Delta(g-r)_0 < \sqrt{0.07^2 + \sigma_g^2 + \sigma_r^2}$, excluding a small number of foreground stars using the high-quality astrometry provided by \Gaia Data Release 3 (DR3;  \citealt{Gaia_Brown_2021A&A...649A...1G, Gaia_Lindegren_2021A&A...649A...2L}; see \secref{propermotions} for more details).
We report the distance modulus that produced the largest number of stars meeting this selection condition. 
\par To define a statistical uncertainty on the distance modulus, we performed a bootstrap resampling (with replacement) of the observed color-magnitude diagram for each candidate 500 times. For each bootstrap iteration, we repeated the procedure described above, finding the distance modulus that maximized the number of stars passing the isochrone selection. 
The standard deviation of the distance moduli across the bootstrap iterations was defined as the statistical error on our best-fit distance modulus. Lastly, we added an systematic uncertainty of 0.1 mag in quadrature to account for the systematic uncertainty associated synthetic isochrone modeling  \citep{2015ApJ...813..109D}, as well as our choice to fix the age and metallicity during this procedure.

\par We performed the procedure described above for DELVE~J1523+2723, DELVE~J1500+0554, and DELVE~J1448+1728, but not for DELVE J1415+3254. DELVE~J1415+3254 was the only candidate for which we detected an unambiguous blue horizontal branch, which provided an excellent anchor for isochrone fitting. Therefore,  we chose to adopt the distance modulus from our initial \ugali fit to the DECam data for this candidate, which was driven by the blue horizontal branch feature; this is likely to be more accurate than a fit to the horizontal branch in the GMOS-N data due to the imperfect relative calibration discussed in \secref{gmosphotometry}. We then proceeded to select the isochrone age and metallicity by eye, as with the other three systems. We assumed a conservative 0.15 mag overall magnitude uncertainty on this horizontal-branch-based distance.
We note that adopting the same distance-fitting procedure as with the other candidates would produce a distance modulus of $(m-M)_0 = 19.86 \pm 0.11$, consistent at the $\sim 1.5\sigma$ level with our quoted result.
\subsubsection{Structural Parameters}
\label{sec:structure}
\par To estimate structural parameters for the candidates with GMOS-N imaging, we first defined a stellar selection from each candidate's calibrated photometric catalog by applying the same cuts on $\chi$ and $sharp$ as for the distance estimation. We then selected all stars consistent with our best-fit isochrone according to $\Delta(g-r)_0 < \sqrt{0.07^2 + \sigma_g^2 + \sigma_r^2}$, where $\sigma_g$ and $\sigma_r$ are the magnitude errors for the $g$ and $r$ bands, which include our GMOS-N photometric error floor of 0.03 mag. Lastly, we once again excluded a small number of foreground stars on the basis of their \Gaia parallaxes and/or proper motions; this time, however, we used the updated distances provided by the analysis in the previous section when performing this filtering.

\par With both the stellar sample and mask defined, we performed a binned Poisson maximum-likelihood fit \citep{2021ApJ...908...18S} to each system assuming a \citet{Plummer:1911} stellar density profile. The free parameters for these fits were the number of observed stars, the X and Y centroid coordinates (in pixels), the extension/semi-major axis length (in pixels), the ellipticity, the position angle (in the image frame), and a background surface density ($\Sigma_b$; in stars per $10^6$ pixels). Unlike \citealt{2021ApJ...908...18S}, we found that it was necessary to model the background surface density in order to account for Milky Way contamination that passed the isochrone filter. The only remaining change in our procedure relative to \citet{2021ApJ...908...18S} was that we multiplied the predicted model counts in each bin by the corresponding value in the binned coverage fraction mask that we constructed above, allowing for accurate comparison with the observed binned counts within the likelihood function.

To validate the results derived from this procedure, we tested our ability to recover structural parameters in an unbiased manner by simulating stellar populations at the catalog level and re-running the structural fit using one of the GMOS-N coverage masks. 
We find that we are able to recover the input structural parameters with appropriately-quantified uncertainties, and we discuss these tests in greater detail in \appref{simtests}.
We also varied the input assumptions for the fit and assessed their impact on each candidate's measured structural properties. These included testing alternate bin sizes, varying the magnitude limit and DAOPHOT parameter cuts used to define the input stellar sample, and masking different subsets of bright stars/galaxies in the full-resolution mask for each system. In general, we found that our results were robust to these choices.

\par Finally, we converted our results from units of pixels to angular units and transformed our position angle measurement from the image frame to the celestial coordinate frame. Specifically, we transformed the posterior probability distributions derived from the MCMC by applying the \code{pixel\_to\_wcs} and \code{position\_angle} functions within \code{astropy} \citep{astropy:2013, astropy:2018}. The angular semi-major axis length of each candidate was converted to a physical quantity with an uncertainty by Monte Carlo sampling from both the posterior distributions for the angular extension and distance. Lastly, the azimuthally-averaged angular (physical) half-light radius was derived from the  angular (physical) semi-major axis length via the relation $r_h = a_h \sqrt{1 - \epsilon}$ ($r_{1/2} = a_{1/2} \sqrt{1 - \epsilon}$).

\subsubsection{Absolute Magnitude}
\label{sec:absmag}
To estimate the absolute magnitude of our candidates from the GMOS-N data, we followed a simulation-based procedure similar to that described by \citet{2008ApJ...684.1075M}. 
We simulated stellar systems consistent with the isochrone parameters and distance of each system assuming a \citet{2003PASP..115..763C} initial mass function. 
We applied a completeness correction to the simulated stars using the completeness functions derived from artificial star tests (\secref{gmosphotometry}) to predicted the number of stars that would be observable brighter that the 30\% completeness limit of our GMOS-N data passing our selections on $\chi$ and $sharp$. 
We adjusted the total stellar mass of each system until the predicted number of observable stars matched the number of member stars returned by our structural parameter fit (\secref{structure}).
This yielded an estimate of the total ``richness'' of the system \citep{2015ApJ...807...50B,2020ApJ...893...47D}, which was then used to derive the absolute magnitude of each system following the method introduced by \citet{2008ApJ...684.1075M}, as implemented by \ugali.
Since most of the luminosity of ultra-faint systems is contributed by the brightest stars, the absolute magnitude estimates have large uncertainties due to the stochastic sampling of the giant branch \citep{2008ApJ...684.1075M}; however, they are reasonably robust to the assumed faint magnitude limit.
\subsection{DECam Data}
\label{sec:decam_properties}
For the candidates for which we obtained follow-up imaging with DECam (DELVE 1921-6047 and DELVE J1057+2852), we performed a maximum-likelihood fit to both the morphological and stellar population properties of each system using \ugali. We modelled the structure of each candidate with a \citet{Plummer:1911} projected density profile, and we simultaneously fit a \citet{Bressan:2012} isochrone to each candidate's observed $g_0,r_0$-band color-magnitude diagram. The eight free parameters for this fit to each candidate were the centroid Right Ascension and Declination ($\alpha_{2000}$ and $\delta_{2000}$, respectively), semi-major axis length ($a_h$), ellipticity ($\epsilon$), and Position Angle east-of-north (P.A.) of the Plummer profile, and the age ($\tau$), metallicity ($Z$), and distance modulus ($(m-M)_0$) of the synthetic isochrone. We sampled this parameter space using the affine-invariant MCMC ensemble sampler \code{emcee} \citep{Foreman-Mackey:2013}, and we report the resulting parameter estimates with their corresponding uncertainties in \tabref{properties}. 

\par Before each fit, we separated stars from background galaxies by applying the selection $0 \leq \var{EXTENDED\_CLASS\_G} \leq 2$, which allowed for a high degree of stellar completeness at the cost of some galaxy contamination (see \citealt{delvedr2}).  In lieu of artificial star tests for the DECam data, we opted to limit the structural fit to sources brighter than the local $10\sigma$ magnitude limit for each candidate, which we estimated within a $30\arcmin$ radius around the discovery centroid of each system. The adopted magnitude limits are provided in \tabref{obsTable}.

\subsection{Proper Motions from Gaia}
\label{sec:propermotions}

We applied proper motion mixture models to further confirm the reality of the systems and to measure the systemic proper motion of the six new satellites following \citet{Pace2019ApJ...875...77P} and \citet{Pace2022arXiv220505699P}. 
Briefly, we modeled the spatial and proper motion distribution with a satellite and Milky Way foreground model. For the satellite distribution, we assumed a \citet{Plummer:1911} density profile for the spatial component and a multi-variate Gaussian distribution for the proper motion distribution. We adopted Gaussian priors for each candidate's extension, position angle, and ellipticity of each system using the \ugali or GMOS-N modeling for the Plummer parameters.
We assumed that the Milky Way foreground was spatially constant over the small area examined and built the proper motion distribution from the distribution of stars at much larger radii with the same target selection. 

We utilized the excellent astrometry associated with {\it Gaia} DR3  \citep{Gaia_Brown_2021A&A...649A...1G, Gaia_Lindegren_2021A&A...649A...2L}, making the following selections to identify stars with good quality astrometry: $ \texttt{ruwe} <1.4,$ $\texttt{astrometric\_excess\_noise\_sig} <2,$ $\texttt{astrometric\_params\_solved} >3,$ and $|C^*| \le 3 \sigma_{C^*}(G)$. We removed foreground stars with a non-zero parallax ($\varpi - 3 \times \sigma_\varpi > 0$) and large proper motions. Specifically, we computed the tangential velocity of each star assuming it lies at the distance of the satellite and removed stars that have velocities significantly larger than the Milky Way escape velocity at that distance ($v_{\rm tangential} - 3 \times \sigma_{v_{\rm tangential}} < v_{\rm escape}$).
We further applied a DECam $g_0$ vs.\ $(g-r)_0$ color-magnitude filter based on an old, metal-poor ($\tau = 13.5 \Gyr$, [Fe/H]$=-2.2$\,dex) PARSEC isochrone, or a similar \Gaia-based $G-G_{RP}$ color-magnitude filter to any stars that lacked $g$ and $r$ photometry in DELVE DR2.  While this analysis could use color--magnitude information as a probabilistic component in the mixture model \citep[e.g.,][]{McConnachie2020AJ....160..124M, Battaglia2022A&A...657A..54B}, we have found that the use of accurate photometry does not require this inclusion to measure systemic proper motions \citep{Pace2022arXiv220505699P}. For a more detailed description of the methodology and data selections presented above, see \citet{Pace2022arXiv220505699P}.
\par We were able to detect a proper motion signal in five of the six candidates reported here, the single exception being DELVE J1500+0554 (see \secref{virgoII}). Compared to many known dwarf galaxies with literature proper motion measurements, the systems we present here have far fewer members in \Gaia \citep[e.g.,][]{Pace2022arXiv220505699P, Battaglia2022A&A...657A..54B}. This is primarily a consequence of the low luminosities of our candidates and not due to their distances. In \secref{review}, we review the properties of the stars that went into our measured proper motions for each candidate, and we provide these stars' \Gaia DR3 source IDs in \appref{appPMs}, \tabref{PMs}.

\section{A Review of the Newly-Discovered Systems}
\label{sec:review}
\begin{figure*}
    \centering
    \includegraphics[width = .8\textwidth]{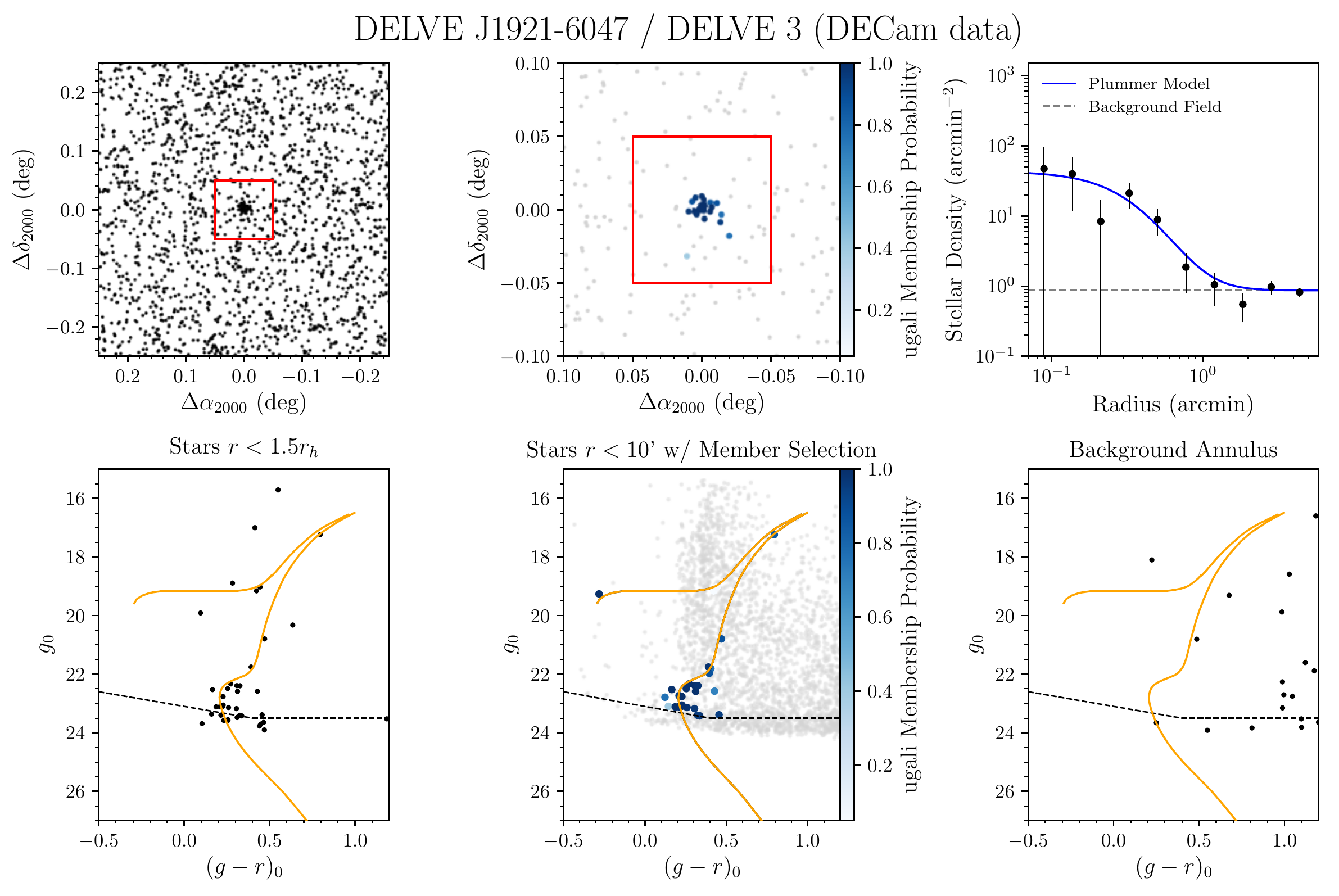}
    \caption{Diagnostic plots for DELVE J1921-6047 (DELVE 3) based on the DECam data. (Top Left) Spatial distribution of all isochrone-filtered stars for a small region centered on the system. (Top Center)  A zoomed-in version of the top left panel, now including only stars above the magnitude limit reported in \tabref{obsTable}. For ease of comparison, the red square denotes the same area as in the left panel. Stars with \ugali membership probabilities greater than five percent ($p_{\ugali} > 0.05$) are shown in blue colors, whereas stars with ($p_{\ugali} < 0.05$) are shown in grey. (Top Right) Radial profile of isochrone-filtered stars centered on DELVE J1921-6047, again including only stars above the magnitude limit. The best-fit azimuthally-averaged Plummer model is shown as a solid blue curve, and the background field density (as estimated within a concentric annulus with an inner radius of $10\arcmin$) is shown as a black dashed line. (Bottom Left) Color--magnitude diagram of stars within a circular selection with a radius $1.5r_h$ for DELVE J1921-6047. A PARSEC isochrone with $\tau = 13.5\Gyr, \rm [Fe/H] = -2.2$~dex is shown as a solid orange curve. The black dashed line denotes the magnitude limit used for our \ugali analysis. (Bottom Center) Color--magnitude diagram of stars within a radius of $r = 10'$, with stars identified as candidate members of DELVE J1921-6047 colored as in the top-center panel.  (Bottom Right) Color--magnitude diagram showing sources within the background annulus used to estimate the field density for the top-right panel. This annulus is equal in area to the circular selection used in the bottom-left panel.
    }
    \label{fig:delve3_6panel}
\end{figure*}

\begin{figure*}
    \centering
    \includegraphics[width = .8\textwidth]{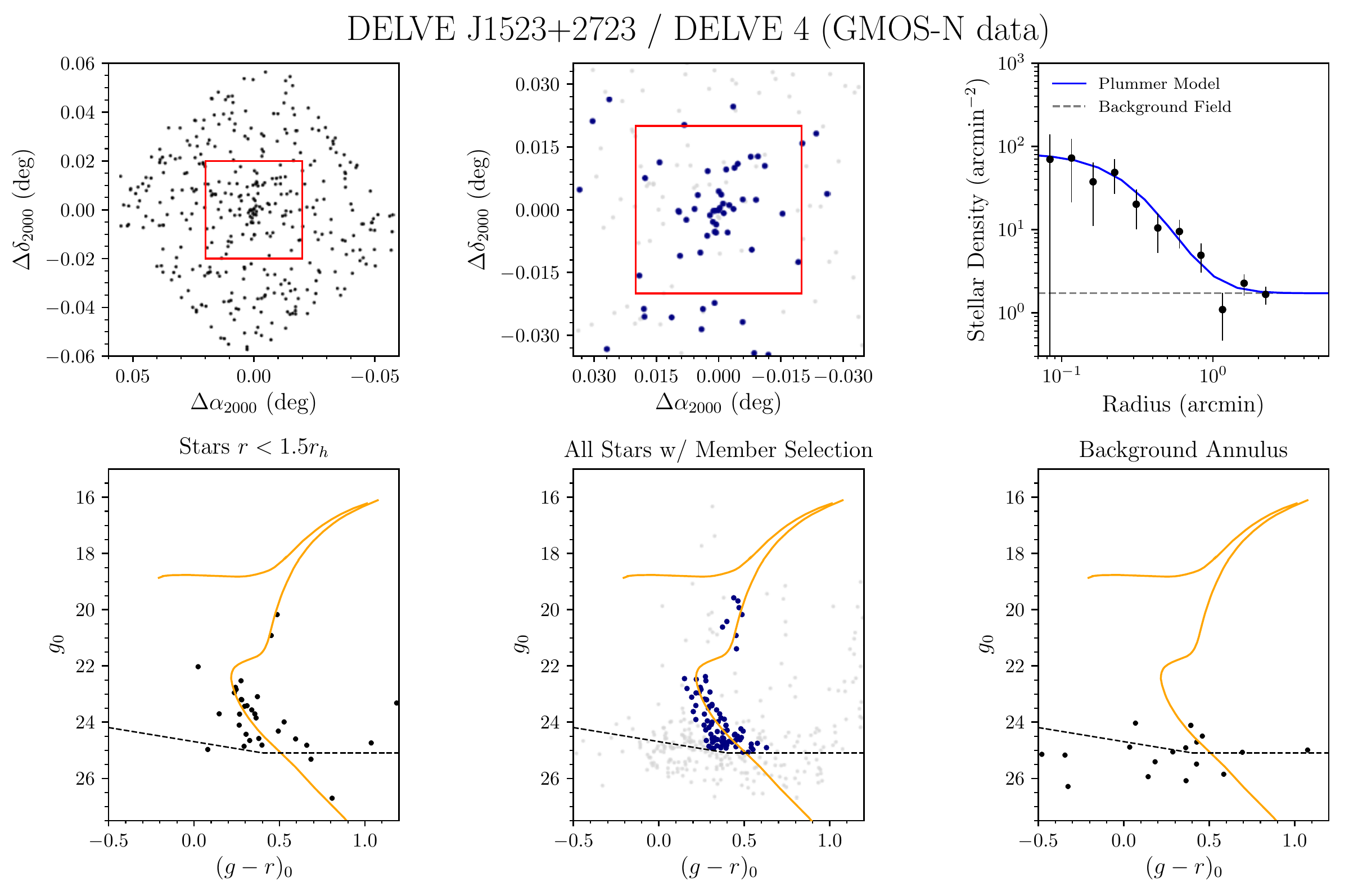}
    \caption{Diagnostic plots for DELVE J1523+2723 (DELVE 4), based on the GMOS-N data. (Top Left) Spatial distribution of all stars in the full GMOS-N FOV; only our quality cut based on $\chi$ and $sharp$ has been applied. (Top Center)  A zoomed-in version of the top left panel, now including only stars above the magnitude limit reported in \tabref{obsTable}. Stars with colors consistent with the isochrone to within 0.07 mag, accounting for photometric errors, are colored in blue and are presumed to be members. (Top Right) Radial profile of isochrone-filtered stars centered on DELVE J1523+2723, again including only stars above the magnitude limit. The best-fit azimuthally-averaged Plummer model is shown as a solid blue curve, and the background field surface density is depicted as a black dashed line. (Bottom Left) Color--magnitude diagram of stars within a circular selection of $1.5r_h$ centered on DELVE~J1523+2723. The best-fit PARSEC isochrone is shown as a solid orange curve. The black dashed line denotes the magnitude limit used for our structural fit. (Bottom Center) Color--magnitude diagram of stars within the GMOS-N FOV, with stars identified as candidate members colored as in the top-center panel. Stars identified as foreground based on their \Gaia parallaxes and/or proper motions are excluded from this member selection, and are shown in grey. (Bottom Right) Color--magnitude diagram for sources in a background annulus with inner radius of $\sim 2.4\arcmin$. This annulus nearly traces the perimeter of the GMOS-N FOV, and is equal in area to the selection used in the bottom-left panel.}
    \label{fig:delve4_6panel}
\end{figure*}

\begin{figure*}
    \centering
    \includegraphics[width = .8\textwidth]{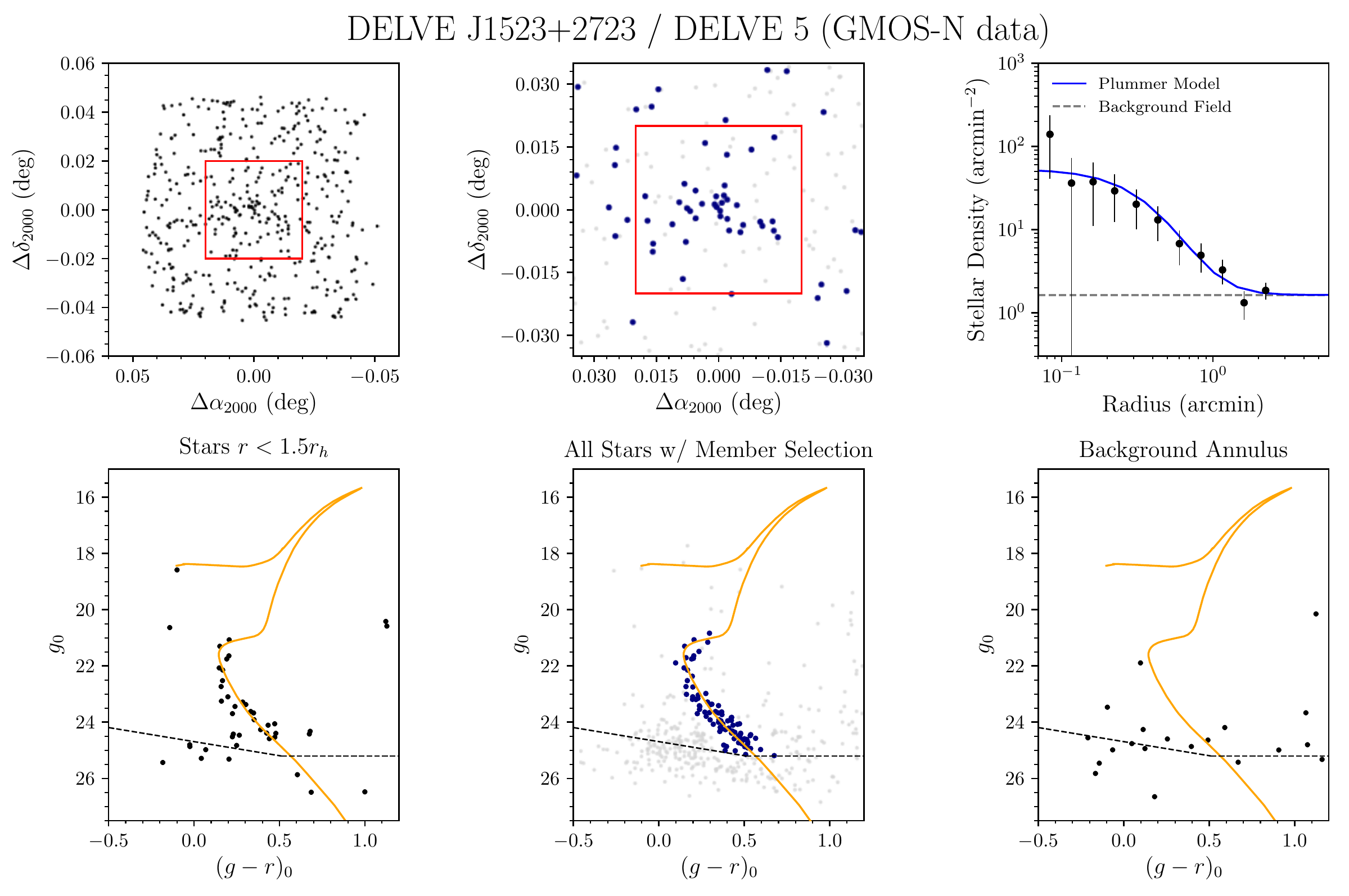}
    \caption{Diagnostic plots for DELVE J1448+1728 (DELVE 5) based on the GMOS-N data. Refer to \figref{delve4_6panel} for a detailed description of each panel. We highlight the unusual morphology of this system, which may suggest that it is not in dynamical equilibrium.}
    \label{fig:delve5_6panel}
\end{figure*}

\begin{figure*}
    \centering
    \includegraphics[width = .8\textwidth]{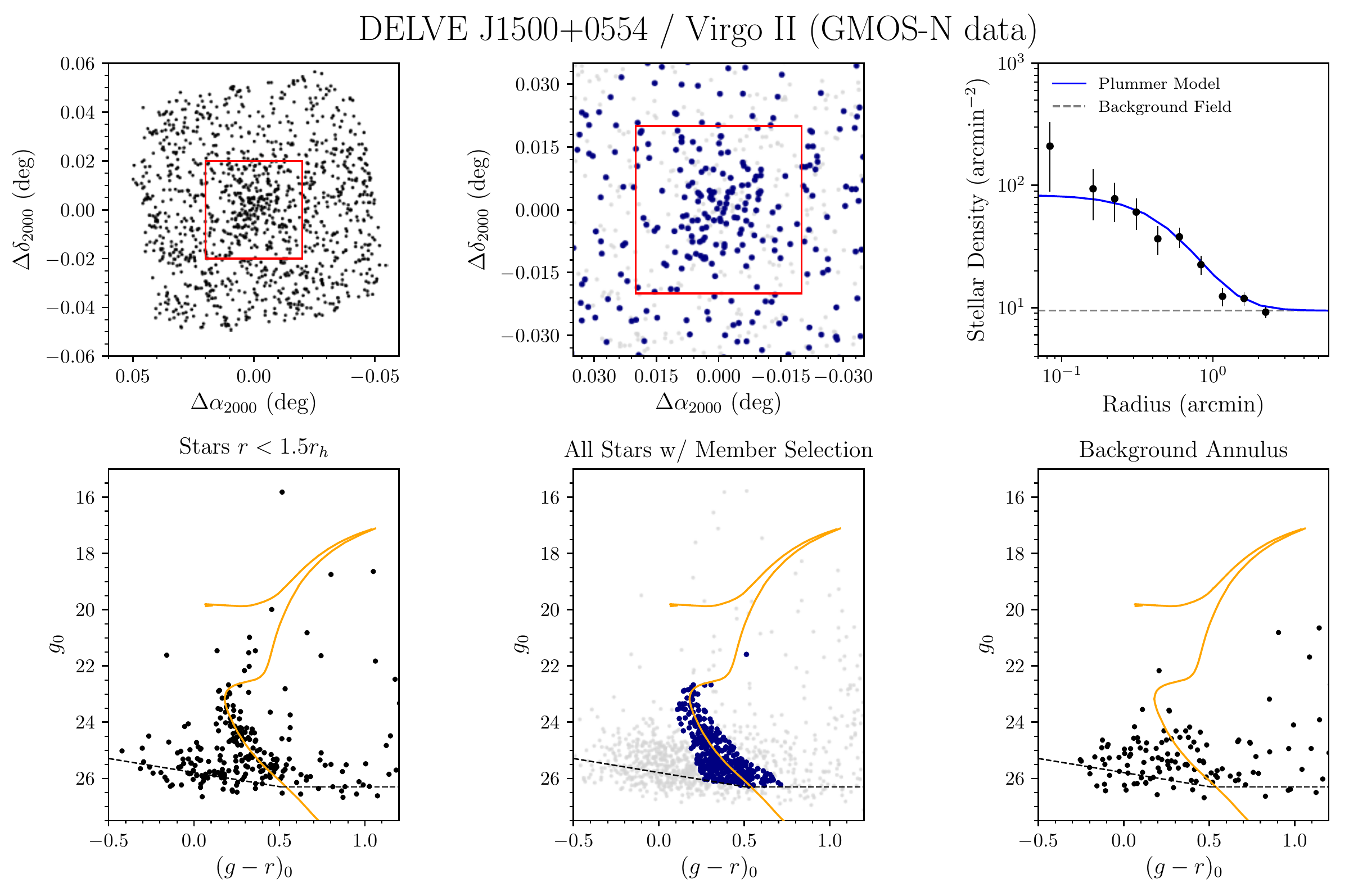}
    \caption{Diagnostic plots for DELVE J1500+0554 (Virgo II) based on the GMOS-N data. Refer to \figref{delve4_6panel} for a detailed description of each panel. }
    \label{fig:virgoII_6panel}
\end{figure*}

\begin{figure*}
    \centering
    \includegraphics[width = .8\textwidth]{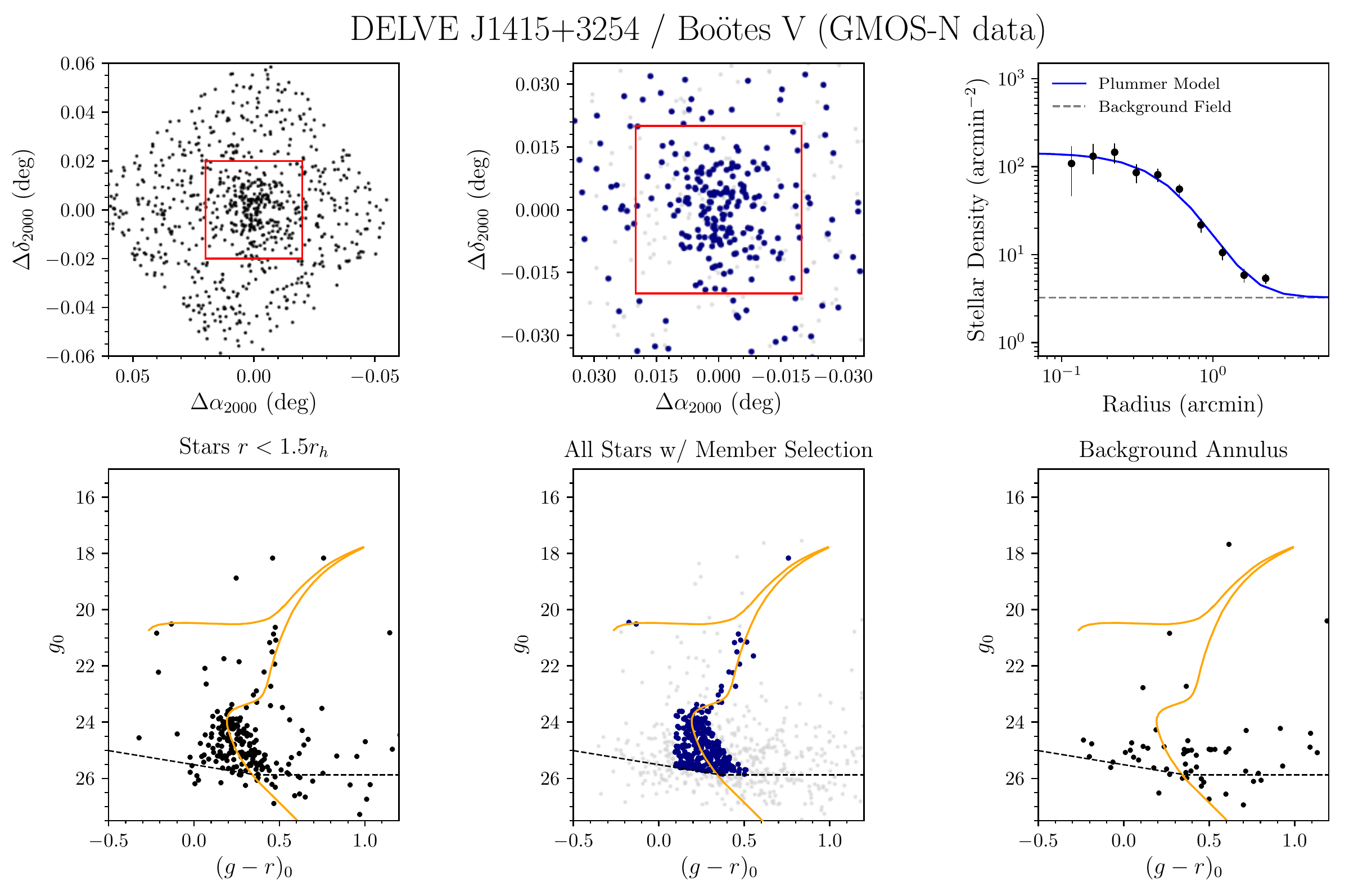}
    \caption{Diagnostic plots for DELVE J1415+3254 (Bo\"{o}tes~V) based on the GMOS-N data. Refer to \figref{delve4_6panel} for a detailed description of each panel. We note that one candidate blue horizontal branch member star falls just outside our isochrone selection.}
    \label{fig:bootesV_6panel}
\end{figure*}

\begin{figure*}
    \centering
    \includegraphics[width = .8\textwidth]{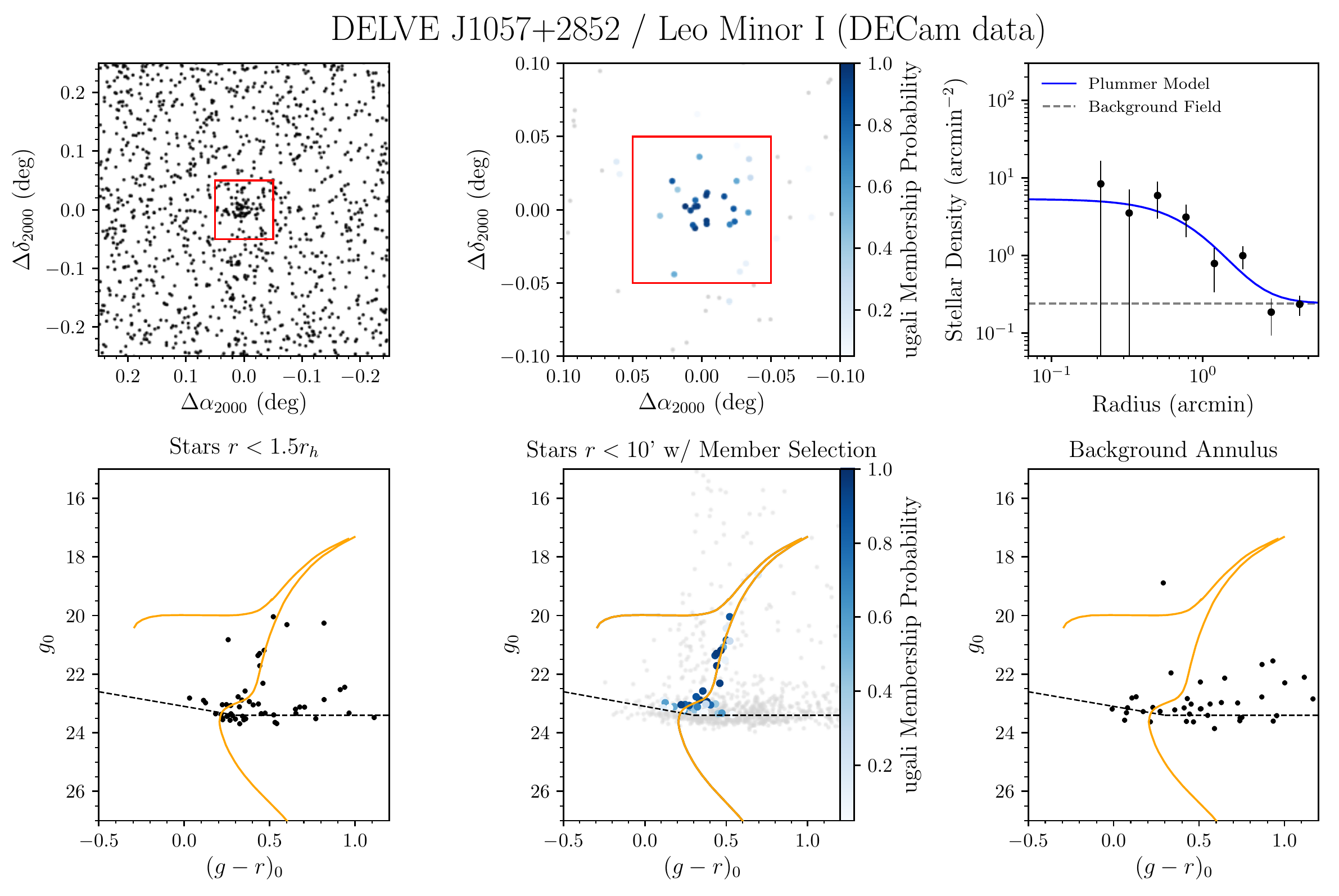}
    \caption{Diagnostic plots for DELVE J1057+2852 (Leo Minor I) based on the DECam data. Refer to \figref{delve3_6panel} for a detailed description of each panel. Although the total number of stars above the magnitude limit is relatively small, the radial stellar density profile appears accurately represented by the best-fit Plummer model.}
    \label{fig:lmi1_6panel}
\end{figure*}

In the following subsections, we review the properties of each of the new candidate systems in greater detail. We begin by discussing the three candidates that we found to be compact before turning to the three candidates that we found to be more extended; we separate these candidates at $r_{1/2} = 15$ pc for convenience. Within this categorization scheme, we order the candidates by decreasing Right Ascension, as in \tabref{obsTable} and \tabref{properties}. Figures \ref{fig:delve3_6panel}--\ref{fig:lmi1_6panel} include the spatial distribution, radial profile, and color--magnitude diagram for stars in the vicinity of each candidate system.

\subsection{Compact Candidates}

\subsubsection{DELVE J1921-6047 (DELVE 3)}

\par Among the candidates presented in this work, DELVE J1921-6047 (\figref{delve3_6panel}) represented the clearest and most significant detection at the time of discovery. It was identified in all four iterations of our \simple search at significances of $\roughly 6.8$--$9.5\sigma$, and is clearly visible as a dense, circular association of blue stars in the color images available through the Legacy Surveys Sky Viewer as part of the Legacy Surveys DR10-early release. Our \ugali fit to deeper DECam data revealed that DELVE J1921-6047 is a compact ($r_{1/2} = 7^{+2}_{-2}$ pc) and low luminosity ($M_V = -1.3^{+0.4}_{-0.6}$ mag) stellar system consistent with an ancient, metal-poor stellar population. The position of DELVE J1921-6047 in the size-luminosity plane of known Milky Way satellites (\figref{popcomparison}) suggests that it is more consistent with the known population of ultra-faint star clusters than the population of known satellite galaxies. Following the convention that ultra-faint star clusters are named after the survey within which they were discovered, we name this system DELVE 3. 

\par We tentatively detect a proper motion signal for DELVE 3 based on two candidate member stars. There is also a nearby candidate blue horizontal branch star with a similar proper motion; however, it is too bright to pass our isochrone selection and therefore holds no weight in our proper motion determination. This star was not identified as a probable member by the \ugali analysis described in \secref{decam_properties}.
\par Although it is impossible to assess the origin of DELVE 3 in the absence of a spectroscopic radial velocity measurement and/or metallicity measurement, the low photometric metallicity and large heliocentric distance ($D_{\odot} = 56^{+2}_{-6}$ kpc) suggested by the best-fit isochrone favor an accreted origin. Importantly, we observe that the system is relatively near to the (on-sky) positions of many ultra-faint dwarf galaxy systems believed to be associated with the Large and Small Magellanic Clouds (hereafter, the LMC and SMC). We further discuss this possible connection between DELVE~3 and the LMC/SMC in \secref{magellanic}.

\subsubsection{DELVE J1523+2723 (DELVE 4)}

\par DELVE J1523+2723 (\figref{delve4_6panel}) was detected with significances of $7.8\sigma$ and $8.1\sigma$ in the $g,r$ searches run with the metal-poor and metal-rich isochrones, respectively. No $i$-band data is available in DELVE DR2 for this system, precluding its discovery in our $g,i$ searches. The system is clearly visible in the Legacy Surveys Sky Viewer as a highly concentrated collection of $\sim 15$ marginally-resolved blue stars that appears comparable in angular size to an adjacent, unresolved spiral galaxy LEDA\,1806653. Despite the clear spatial signature for this system at the image level, the DECam photometry was insufficient to constrain its properties or reveal a clear signature in color--magnitude space.
\par With the improved depth and resolution of our GMOS-N data, we were able to resolve many more stars in this system, revealing a clear main sequence in our color--magnitude diagram consistent with an ancient stellar population at a heliocentric distance of $D_{\odot}$~$\sim 45^{+4}_{-4}$~kpc. The stellar population of DELVE J1523+2723 is adequately described by a PARSEC isochrone with $\rm [Fe/H] \sim -1.9$ dex, but this measurement may be unreliable due to the dearth of red giant branch stars. Our structural fit to this system affirms that it is very compact ($r_{1/2} = 5^{+1}_{-1}$ pc) and faint ($M_V = -0.2^{+0.5}_{-0.8}$ mag). Based on these properties, we conclude that this system is most likely an ultra-faint star cluster, and we therefore name it DELVE 4.
\par We detect a proper motion signal in DELVE~4 based on six faint red giant branch stars. Although two of these stars are located at a fairly large distance relative to the center of the system ($\gtrsim 3r_h$), they have similar proper motions to the four stars near the system's centroid. The proper motion of DELVE~4 is discrepant with the Milky Way foreground proper motion in the region of sky near DELVE~4, and we conclude that we have confidently detected the systemic proper motion of DELVE~4. Speculatively, the large ellipticity ($\epsilon\sim0.5$) and the presence of the two distant candidate members may suggest that DELVE~4 is undergoing tidal disruption. To confirm this scenario, radial velocity measurements for the distant members are required. 

\subsubsection{DELVE J1448+1728 (DELVE 5)}
We identified DELVE J1448+1728 (\figref{delve5_6panel}) at $\roughly 7.1\sigma$ in the $g,r$ iterations of our \simple search; again, no $i$-band data is available in DELVE DR2 for this system.  Our initial diagnostic plots revealed two possible blue horizontal branch star candidates, and inspection of the color images in the Legacy Surveys Sky Viewer revealed a compact clustering of blue sources. However, the seemingly unusual distribution of these sources in the system was difficult to interpret. Furthermore, the Dark Energy Spectroscopic Instrument targeting overlay in the Sky Viewer incorrectly suggested that the bulk of sources in this system were either quasars or emission line galaxies (see \citealt{2022arXiv220808511C} and \citealt{2022arXiv220808513R} for details on the target selection and expected contamination rates for each class of objects).
\par Despite these apparent anomalies, the GMOS-N data for DELVE J1448+1728 clearly reveals an extended main sequence consistent with an old, metal-poor stellar population. The distribution of isochrone-selected members in the GMOS-N FOV remains somewhat unusual, potentially suggesting that the system is either very diffuse and/or that contaminants may be obfuscating our view of the system. Our structural fit suggests that this system is exceedingly faint $M_V = +0.4^{+0.4}_{-0.9}$, with a highly elliptical morphology ($\epsilon = 0.6^{+0.1}_{-0.2}$, $r_{1/2} = 5^{+1}_{-1}$~pc).  One explanation for the sum total of these features is that the system is an ultra-faint star cluster that is undergoing tidal disruption, or simply evaporating due to N-body relaxation \citep[e.g.,][]{1997A&ARv...8....1M}. This would explain the apparently low luminosity (induced via mass loss) as well as the unusual spatial extension.  
\par Ascertaining the true nature of DELVE J1448+1728 will likely require deep imaging over a wider field of view. Nonetheless, we observe a clear similarity between the properties of this system and those of the dissolving ultra-faint star cluster Kim 1. Specifically, Kim 1 features a similar lack of central concentration, moderately high ellipticity ($\epsilon = 0.4$), small size ($r_{1/2} = 6.9$~pc), and low luminosity ($M_V = 0.3$)  \citep{2015ApJ...799...73K}. On the basis of this similarity, we presume that DELVE J1448+1728 is also an ultra-faint star cluster out of dynamical equilibrium, and we name the system DELVE~5.

\par We detect a possible proper motion signal for DELVE~5 based on three stars located at relatively large angular distances from the center of the system; these include two blue horizontal branch stars and a red giant branch star at a similar magnitude. Oddly, we found no fainter members. It is possible that the signal we have identified is not the signal of DELVE 5, and it may instead represent the motion of another structure in the Milky Way halo. Radial velocities of these three stars would enable confirmation of our measurement, and would allow for a more detailed assessment of the system's dynamical state.

\subsection{Extended Candidates}

\subsubsection{DELVE J1500+0554 (Virgo II)}
\label{sec:virgoII}
DELVE J1500+0554 (\figref{virgoII_6panel}) was identified in all four iterations of our \simple search with detection significances ranging from $5.6\sigma$ to $7.9\sigma$. It was visually confirmed in the Legacy Surveys Sky Viewer as a concentration of faint blue stars spread across a spatial region that was at least $2\arcmin \times 2\arcmin$ in size.  This system is located in the constellation Virgo and falls $\sim 3 \degree$ above the northern boundary of the Stripe 82 subsection covered by the second data release of HSC SSP (HSC SSP DR2); this explains why this candidate was not discovered in tandem with the Virgo I ultra-faint dwarf \citep{2016ApJ...832...21H}.
\par
\begin{figure*}
    \centering
    \includegraphics[scale=.8]{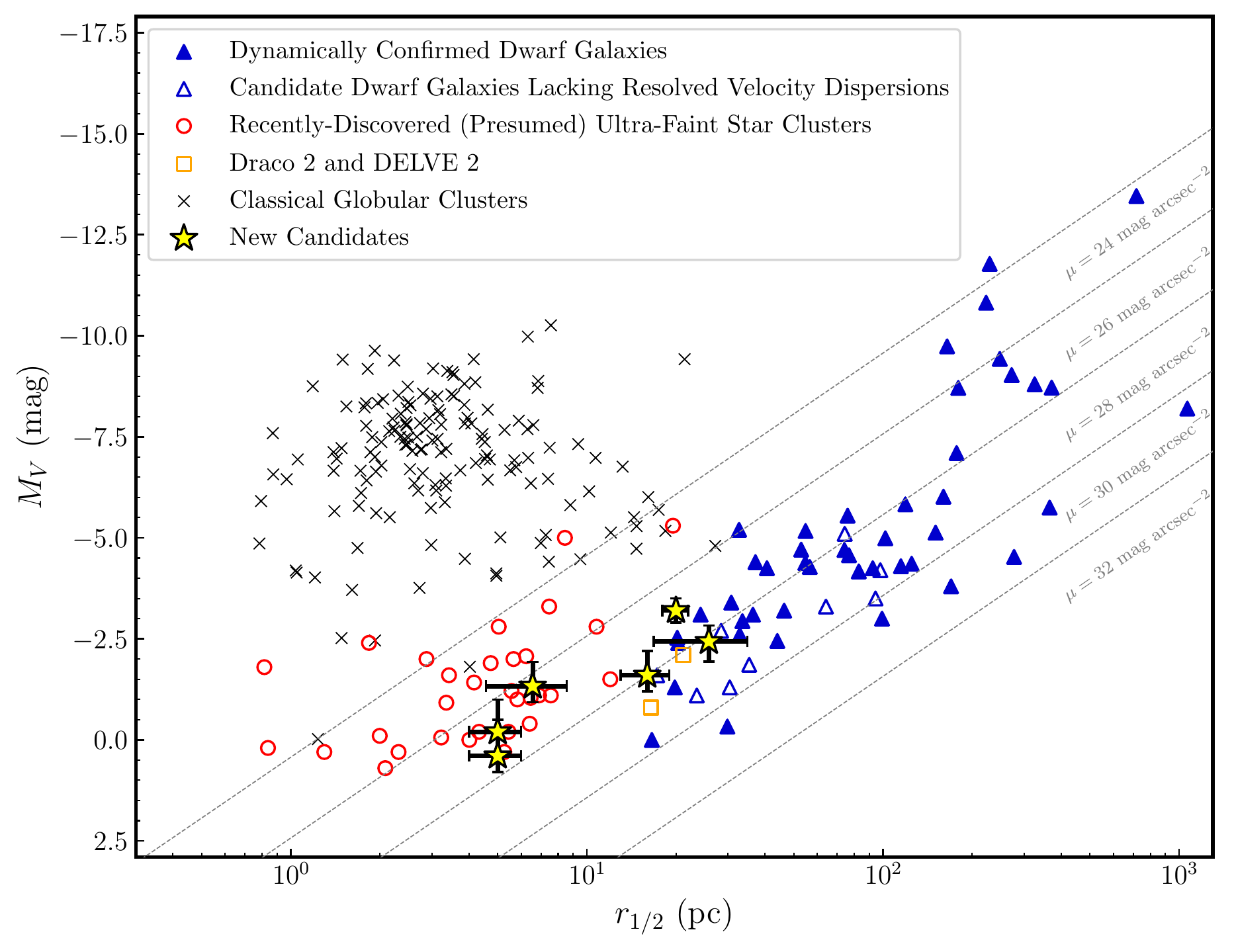}
    \caption{Absolute $V$-band magnitude $(M_V)$ vs.\ azimuthally-averaged physical half-light radius ($r_{1/2}$) for various different populations of Milky Way satellites. Dynamically confirmed dwarf galaxies (i.e., those with resolved velocity dispersions) are shown as solid blue triangles, and candidate dwarf galaxies (with unresolved dispersions) are shown as open blue triangles. ``Classical'' Milky Way globular clusters from \citet[][]{1996AJ....112.1487H}, 2010 edition are shown as black crosses. Recently-discovered systems presumed to be ultra-faint star clusters are shown as open red circles. The systems DELVE 2 \citep{2021ApJ...920L..44C} and Draco II \citep{2015ApJ...813...44L}, which have as-yet unknown classifications, are included as orange squares. For a complete reference list, see Appendix A of \citet{2022arXiv220311788C}. Lastly, the six new candidates presented in this work are shown as yellow stars. From left to right (i.e., in order of increasing $r_{1/2}$), these are DELVE 4 and DELVE 5 (at the same $r_{1/2}$, where DELVE~4 is brighter), DELVE 3, Virgo II, Bo\"{o}tes~V, and Leo Minor I.}
    \label{fig:popcomparison}
\end{figure*}

 With our GMOS-N imaging, we confidently resolve stars three magnitudes below the main sequence turnoff, revealing a large sample of possible member stars consistent with the best-fit old, metal-poor PARSEC isochrone. Our structural fit reveals that DELVE J1500+0554 is a relatively small, ultra-faint system ($r_{1/2} = 16^{+3}_{-3}$ pc; $M_V = -1.6^{+0.4}_{-0.6}$) at a distance of $D_{\odot} = 72^{+8}_{-7}$ kpc. This physical size places the system within the so-called ``trough of uncertainty'' \citep{2018ApJ...852...68C} in the $M_V - r_h$ plane within which the populations of ultra-faint star clusters and ultra-faint dwarf galaxies overlap (see \figref{popcomparison}). Its closest analogs are Draco II, Cetus II, and Triangulum II, which have comparable half-light radii ($r_{1/2} = [16, 17, 17]$ pc, respectively) and luminosities $M_V = [-0.8, 0.0, -1.6]$ \citep{2018MNRAS.480.2609L,2015ApJ...813..109D,2017AJ....154..267C}. All three of these known systems have proven difficult to classify. Neither Draco II nor Triangulum II have resolved velocity or metallicity dispersions, and thus have required more oblique arguments toward their classification. Toward this end, \citet{2022MNRAS.510.3531B} searched for signatures of mass segregation in each of these systems. They found clear evidence for mass segregation in Draco II, and conclude that the system is a star cluster; by contrast, they find that Triangulum II is unsegregated, favoring a dwarf galaxy classification. \citet{2019ApJ...870...83J} also argue that Triangulum II is a dwarf galaxy on the basis of its low neutron-capture element abundances measured through high-resolution spectroscopy. The reality of Cetus II has been questioned by \citet{2018ApJ...857...70C}, and we do not consider it further here.
\par As the above discussion reveals,  the morphological properties of known ultra-faint stellar systems with $r_{1/2} \sim 16-17$ pc are insufficient to reach a firm determination about their classifications. Indeed, the cases of Draco II and Triangulum II convey that systems of this size could reasonably be either star clusters or dwarf galaxies in the absence of additional information. This effectively precludes us from making a high-confidence classification for DELVE J1500+0554.
However, given the close similarity of \textit{both} the absolute magnitude and half-light radius between DELVE J1500+0554 and Triangulum II, we believe that DELVE J1500+0554 is more likely to be an ultra-faint dwarf galaxy. We therefore adopt the name Virgo II, but emphasize that spectroscopic follow-up is necessary to reveal its true nature.

\par We do not detect a \Gaia proper motion signal for Virgo~II.  There are only two stars that pass the data quality and color-magnitude filtering within $4\arcmin$ of the Virgo~II center. These two stars have discrepant proper motions. With radial velocity information, it will likely be possible to distinguish which of these stars (if either) are members of the system and thereby constrain the proper motion. At this time, however, we do not report a proper motion measurement for this system.

\subsubsection{DELVE J1415+3254 (Bo\"{o}tes~V)}

DELVE J1415+3254 (\figref{bootesV_6panel}) was detected in both $g,r$ iterations of the search, but only barely surpassed our $5.5\sigma$ detection threshold for the iteration that used a metal-rich isochrone. Despite its low detection significance, we flagged this candidate for further investigation due to the apparent presence of three blue horizontal branch stars in our diagnostic plots from \simple. We also identified that multiple stars in the immediate vicinity displayed self-consistent proper motions, supporting the reality of the system.  With the deeper GMOS-N data, we were able to robustly constrain DELVE J1415+3254's properties, finding that it is an intermediate-size ($r_{1/2} = 20^{+2}_{-2}$ pc), relatively circular ($\epsilon = 0.2^{+0.1}_{-0.1}$), and low-luminosity ($M_V = -3.2^{+0.3}_{-0.3}$) system  in the outer Milky Way halo ($D_{\odot} = 102^{+7}_{-7}$ kpc). Its stellar population is well described by the most metal-poor isochrone in our PARSEC grid ($\rm [Fe/H] = -2.2$~dex).
\par  Like Virgo II, DELVE J1415+3254's absolute magnitude and half-light radius place it in a region of parameter space where the populations of ultra-faint dwarf galaxies and ultra-faint star clusters overlap. This makes the system challenging to classify based on morphology alone. A physical size of 20\,pc is not large enough to unambiguously suggest a dwarf galaxy classification for the system, especially given the measurement uncertainties at play. Ultra-faint systems with similar morphological and stellar population properties to DELVE J1500+0554 include Willman 1, Carina III, Tucana V, Pictor I, and Phoenix II, which have $r_{1/2} = [20, 20, 24, 24, 28]$ pc and $M_V = [-2.53, -2.4,-1.1, -3.1, -2.7]$, respectively \citep{2018ApJ...860...66M,2018MNRAS.475.5085T,2020ApJ...892..137S,2018ApJ...863...25M}. Despite its irregular kinematic distribution, \citet{Willman2011AJ....142..128W} argue that Willman 1 is a dwarf galaxy on the basis of its metallicity spread. \citet{2018ApJ...857..145L} and \citet{2020ApJ...889...27J} both argue Carina III is a dwarf galaxy based on a large metallicity spread between member stars, and the latter work further supports this argument by establishing that the system displays low neutron-capture element abundances and no light element anticorrelatations. \citet{2019A&A...623A.129F} identify a large velocity dispersion for Phoenix~II, robustly establishing its nature as a dwarf galaxy. No published spectroscopic measurements exist for Tucana~V or Pictor~I.
\par Based on the likelihood that three of the five systems with similar morphological properties to DELVE J1415+3254 are dwarf galaxies, we conclude that DELVE J1415+3254 is also most likely a dwarf galaxy. Accordingly, we name the system Bo\"{o}tes~V following the convention that dwarf galaxy candidates are named after the constellation within which they reside; this name matches that given by the contemporaneous work by \citet{Smith:2022}.

\par Regardless of its classification,  Bo\"{o}tes~V features the most confident \Gaia proper motion signal of the six systems analyzed. It features a bright red giant branch star, two horizontal branch stars, and several fainter red giant branch stars with self-consistent proper motions, although the majority of the proper motion measurement precision is attributable to the brightest star alone. 
\par Despite the absence of a radial velocity measurement for Bo\"{o}tes~V, we briefly explored its orbital properties using the \texttt{gala} package, assuming the package's default \texttt{MilkyWayPotential} Galactic potential \citep{gala}. 
We scanned through a range of heliocentric line-of-sight velocities for which the system is most likely bound to the Milky Way ($-300\kms < v_{\rm los} < 600 \kms$) in steps of $5\kms$. At each velocity, we integrated the orbits for 10~Gyr and vary the distance and proper motion according to their observational errors.
\par
\begin{figure}
    \centering
    \includegraphics[width=\columnwidth]{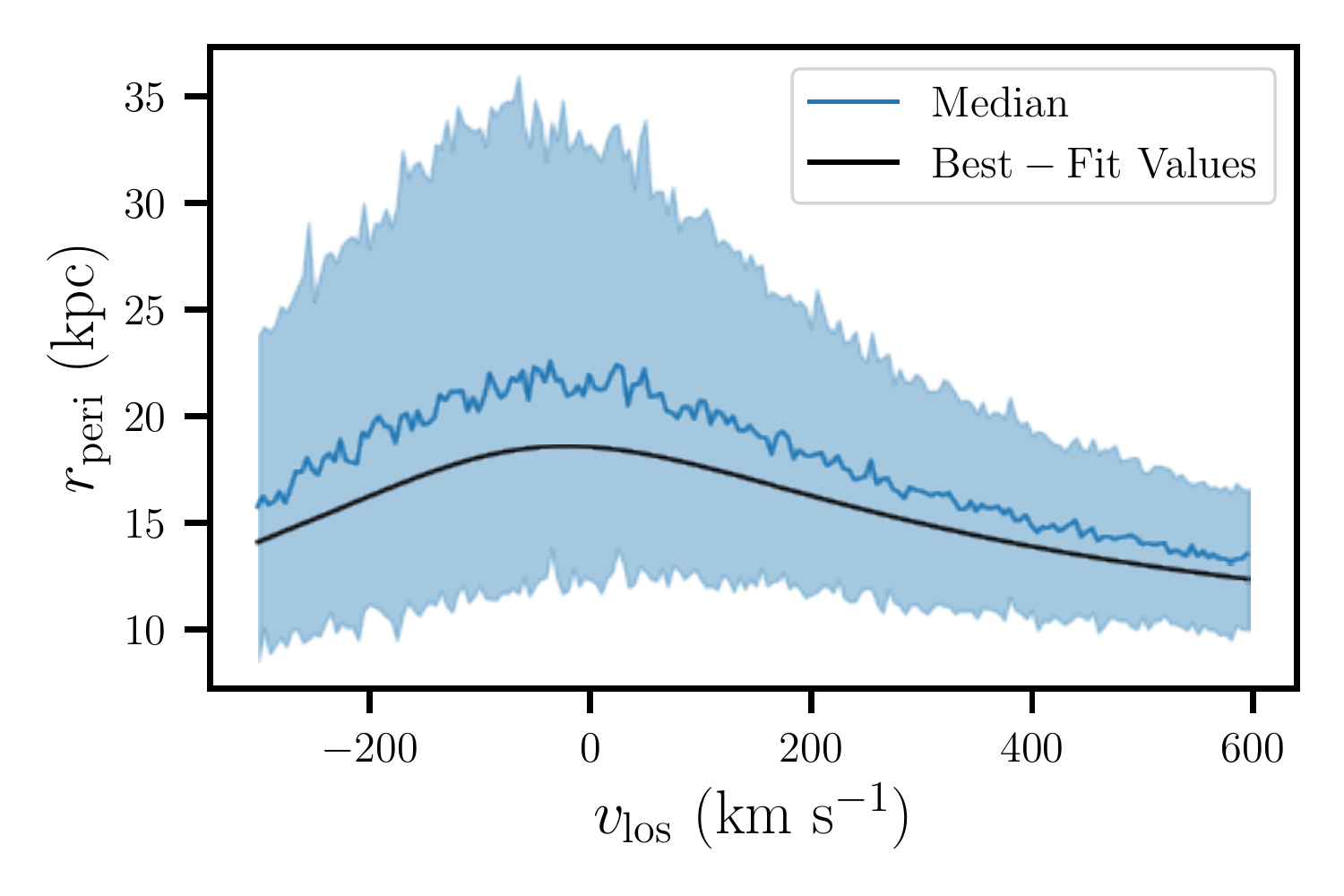}
    \caption{Predicted orbital pericenter ($r_{\rm peri}$) across a range of possible line-of-sight velocities ($v_{\rm los}$) for Bo\"{o}tes~V. The black line uses the median values for the proper motion and distance, whereas the blue line and bands are from MCMC sampling of the proper motion and distance errors.}
    \label{fig:bootes_5_orbit}
\end{figure}

 We found that  Bo\"{o}tes~V is on a polar orbit with a minimum predicted apocenter at a $v_{\rm los}\sim-80~\kms$. Interestingly, the system's proper motion implies a small pericenter ($r_{\rm peri} \lesssim 15-20$ kpc) across a wide range of possible radial velocities (see \figref{bootes_5_orbit}).
Such a close pericentric passage could result in tidal elongation or stripping of the system, as has been observed in a small number of known ultra-faint systems with small pericenters \citep[e.g.,][]{Li:2018}. Although we found no evidence for any morphological features suggestive of tidal disruption, this unique property of  Bo\"{o}tes~V strongly merits further study. Specifically, measurement of the system's radial velocity would allow confirmation of the low pericenter, and comparably deep imaging over a wider FOV would enable searches for distant member stars and extended structure around the system \citep[e.g.,][]{2007ApJ...668L..43C,2010ApJ...718..530S, 2012ApJ...756...79S,  2019ApJ...885...53M, 2020ApJ...902..106M}.

\subsubsection{DELVE J1057+2852 (Leo Minor I)}
\label{sec:leo_minor_1}

Like Bo\"{o}tes~V, DELVE J1057+2852 (\figref{lmi1_6panel}) also lies near the northern edge of the sky accessible to DECam. It was detected in all four iterations of our search at significances of $6.1\sigma$ to $7.3\sigma$,  despite the presence of CCD gaps in the surrounding field within DELVE DR2 (\figref{mapfigure}). The system is visible in the Legacy Surveys Sky Viewer as a cloud of marginally-resolved blue stars extending at least two arcminutes in diameter (similar to Virgo II).
\par Despite our additional imaging of this system with DECam, the data available for DELVE J1057+2852 are relatively shallow for characterizing the structure of such a faint system at this distance ($D_{\odot} = 82^{+4}_{-7}$ kpc). In particular, the small sample of available member stars, combined with the lack of blue horizontal branch stars or a fully-resolved main sequence turnoff, resulted in considerable uncertainties on the best-fit distance and size of this system. These issues were likely compounded by the presence of stars from the Sagittarius stream in the foreground (see \secref{sagittarius}), as well as the existence of a moderately bright star positioned near the system centroid. In an attempt to better constrain its properties, we explored nearly two dozen different combinations of magnitude limits for the $g,r$ bands, as well as different star/galaxy separation criteria. This resulted in a range of angular half-light radii of $r_h \sim 0.6' - 1.7'$, with shallower magnitude limits generally resulting in smaller angular sizes (and vice versa). The parameters we report come from our nominal choice to limit sources for the fit to stars above the $10\sigma$ magnitude limit, but we caution that deeper imaging is essential for accurate measurements for this candidate. Nonetheless, the large uncertainty on our measurement of $r_h$ encompasses much of this range.
\par Accepting the caveats above, DELVE J1057+2852 is the most extended and second most luminous system we report in this work (after Bo\"{o}tes~V), with an azimuthally-averaged half-light radius of $r_{1/2} = 26^{+9}_{-9}$ pc and an absolute magnitude of $M_V = -2.4^{+0.5}_{-0.4}$ mag. At face value, the one-sigma confidence interval for the half-light radius spans from systems like Draco II, Triangulum II, and Bo\"{o}tes~V (on the small end) to systems like Horologium I, Sagittarius II, Horologium II, Bo\"{o}tes II, and Segue 2 (on the large end). Among the latter set of systems, Horologium I, Bo\"{o}tes II, and Segue 2 are accepted to be dwarf galaxies on the basis of their velocity and/or metallicity dispersions.  Sagittarius II displays a small velocity dispersion but is most likely a star cluster \citep{2020MNRAS.491..356L, 2021MNRAS.503.2754L}. However, despite its extended size, Sagittarius II is not closely analogous to DELVE J1057+2852 due to its significantly brighter absolute magnitude ($M_V = -5.2$). On the other hand, Horologium II lacks a sufficiently large spectroscopic sample to reach any firm conclusions about its classification \citep{2019A&A...623A.129F}.  
\par In pursuit of a direct classification for DELVE J1057+2852, we obtained medium-resolution spectroscopy for this candidate with the DEep Imaging Multi-Object Spectrograph (DEIMOS; \citealt{2003SPIE.4841.1657F}) instrument on the Keck II telescope. 
These data will be presented in detail in future work; however, a preliminary analysis revealed a clustering of stars with radial velocities of $0$--$6~\text{km s}^{-1}$. Five of these stars are very metal poor, with calcium-triplet-derived metallicities of $\rm [Fe/H] \sim -2.5$ dex (using the calibration from \citealt{2013MNRAS.434.1681C}). These stars display hints of a small, but measurable, spread in velocities, suggestive of a dwarf galaxy classification for this system, although this result should be interpreted cautiously given its preliminary nature and the small number of observed stars. Deeper imaging, as well as a more complete spectroscopic dataset, will undoubtedly aid us in elucidating the nature of this system. Nonetheless, we believe that the totality of the evidence presented above favors a dwarf galaxy classification for this candidate. We therefore name this system Leo Minor I.
\par Lastly, we detected a proper motion signal for Leo Minor I in \Gaia through our mixture modelling approach; this signal is derived from three faint red giant branch stars. These three stars have similar radial velocities measured from the DEIMOS spectroscopy, and two of these stars have similar metallicities, giving us confidence in the accuracy of this proper motion measurement despite the small sample size.
\begin{figure*}
    \centering
    \includegraphics[width = \textwidth]{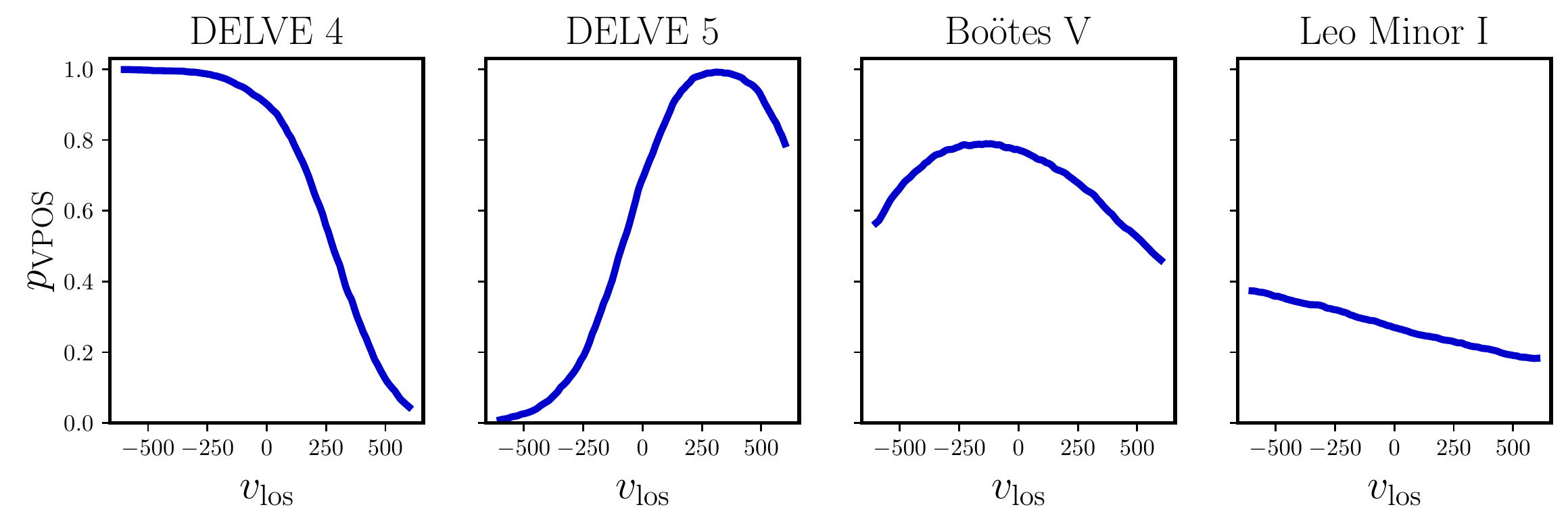}
    \caption{Predicted VPOS membership probability $p_{\rm VPOS}$ as a function of radial velocity for the five candidates lying at $\theta_\text{pred} < 36.9 \degree$. We exclude DELVE 3 because its spatial position falls well outside the VPOS plane, and we exclude Virgo II because we were unable to measure its proper motion. Overall, we find that DELVE 4, DELVE 5, and Bo\"{o}tes~V are promising VPOS member candidates, and that VPOS membership for Leo Minor I is disfavored but not ruled out. We caution that not all radial velocities are equally probable, and that radial velocities at the extremes of the ranges presented here may correspond to orbits that are unbound.}
    \label{fig:vpos}
\end{figure*}
\section{Association with Local Group Structures}
\label{sec:association}
\par We investigated whether the six stellar systems reported here may be associated with several Local Group structures, including the Sagittarius dwarf spheroidal galaxy, the Vast Polar Structure, and the Magellanic Clouds.

\subsection{The Sagittarius Dwarf Spheroidal}
\label{sec:sagittarius}
\par The Sagittarius dwarf spheroidal galaxy (Sgr), discovered by \citet{1994Natur.370..194I}, represents one of the most significant stellar substructures in the Milky Way environment, featuring both a dense core positioned near the Galactic plane as well as an extended tidal stream stretching across the entire sky. In recent years, considerable effort has been invested into attempts to identify stellar systems that may have accreted along with Sgr, including both globular clusters and dwarf galaxies \citep[e.g.,][]{1995AJ....109.2533D,2003AJ....125..188B,2010MNRAS.404.1203F,2010ApJ...718.1128L,2017MNRAS.468...97L,2021A&A...654A..23G}. 
\par Two of our dwarf galaxy candidates, Leo Minor I and Virgo II, lie at low latitudes relative to the Sgr stream (Sgr stream latitude of $|\beta| < 11\degree$, in the coordinate system of \citealt{2010ApJ...714..229L}), raising the possibility of a similar association for these systems. To investigate whether these candidates could plausibly have been accreted with Sgr, we turned to the Sgr models of \citet{2021MNRAS.501.2279V}, who performed tailored N-body simulations of the orbit of Sgr around the Milky Way in the presence of the LMC. We selected all simulated star particles within a radius of $5\degree$ centered on Leo Minor I and Virgo II, and considered whether the distances and velocities of the particles matched those of the two candidates, where possible.
\par Near Leo Minor I, we found two groups of Sgr stars: one group with heliocentric distances $D_{\odot} \sim 10-30$ kpc and line-of-sight velocities $ -165 \text{ km s}^{-1}$ to $-20 \text{ km s}^{-1}$, and another group with $D_{\odot} \sim 45-75$ kpc and $v_{\rm los} \sim 160 \text{ km s}^{-1}$ to $-200\text{ km s}^{-1}$. Leo Minor I is consistent with membership in the latter group on the basis of its heliocentric distance ($D_\odot = 82^{+4}_{-7}$\,kpc); however, its line-of-sight velocity (tentatively, $\sim 4 \text{ km s}^{-1})$ strongly disfavors an association with either group. At the location of Virgo II, we observe three separate groups of Sgr stars in the $D_{\odot} - v_{\rm los}$ plane. The farthest of these groups lies at $D_{\odot} \sim 45-60$ kpc, to be compared with $D_{\odot} = 72^{+8}_{-7}$ kpc for the candidate dwarf. This difference in distance weakly disfavors an association between Virgo~II and Sgr, although a proper motion and/or radial velocity measurement for the former would allow for a more informative comparison.

\subsection{The Vast Polar Structure}
We considered whether any of the dwarf galaxy candidates presented here are plausibly associated with the Vast Polar Structure (VPOS) of the Milky Way \citep{Pawlowski:2012}. A substantial fraction of the observed Milky Way satellite galaxies have aligned orbital planes oriented nearly perpendicular to the Milky Way's stellar disk \citep{2013MNRAS.435.2116P, 2020MNRAS.491.3042P}, extending to the faintest galaxies \citep{2018A&A...619A.103F} with updated \Gaia EDR3 proper motions \citep{Li:2021}.

Specifically, we considered two quantities to assess VPOS membership: the minimum possible angle, $\theta_\text{pred}$, between the VPOS normal and the satellite's orbital pole based on spatial information alone \citep{2013MNRAS.435.2116P}, and the probability, $p_\text{VPOS}$, that the satellite's orbital pole lies within some angular tolerance, $\theta_\text{in,VPOS}$, of the VPOS normal after incorporating proper motions and measurement uncertainties. Since all but one of our candidates do not have measured line-of-sight velocities, we calculated $p_\text{VPOS}$ for a range of assumed velocities from $-600 \kms < v_{\rm los} < +600 \kms$ (see \figref{vpos}). We adopt the same methodology and VPOS parameters as prior work \citep{2018A&A...619A.103F, Li:2021}, namely the assumed VPOS normal $(l_\text{MW}, b_\text{MW}) = (169.3\degree, -2.8\degree)$  and angular tolerance $\theta_\text{in,VPOS} = 36.87\degree$, corresponding to circles that encompass 10\% of the sky. 
\par We display the results of this analysis in \figref{vpos}. In brief, we found that the spatial positions and proper motion measurements favor membership for Bo\"{o}tes~V ($0.5 < p_\text{VPOS} < 0.8$), but mildly disfavors membership for Leo Minor I ($p_\text{VPOS} \sim 0.27$ at the tentative radial velocity of $4 \text{ km s}^{-1}$). We could not explore the same possibility for Virgo II given the lack of proper motion or radial velocity measurement. However, we do observe that its spatial position suggests a minimum angle $\theta_\text{pred} \sim 33\degree$. This is less than the tolerance adopted by \citet{2018A&A...619A.103F} and \citet{Li:2021}, suggesting that an association may be possible.
\par Although the orbits of globular clusters and stellar streams do not preferentially align with the VPOS \citep{Riley:2020}, we also explored the possibility VPOS membership for our ultra-faint star cluster candidates (DELVE 3--5) in the same manner. We found that DELVE 3's membership is ruled out based on its spatial position alone, while DELVE 4 and DELVE 5 are favored or heavily favored to be VPOS members ($0.5 < p_\text{VPOS} < 1$), although the degree depends on the radial velocity.

\subsection{The Magellanic Clouds}
\label{sec:magellanic}

A growing body of observational evidence suggests that many of the known Milky Way satellite galaxies were accreted along with the Magellanic Clouds as they fell into the Milky Way \citep[e.g.,][]{2015ApJ...807...50B, 2015ApJ...805..130K, 2017MNRAS.465.1879S, 2018ApJ...867...19K,2020ApJ...893..121P,2019A&A...623A.129F,2020MNRAS.495.2554E}.  Although five of our six candidates are unlikely to be associated with the LMC/SMC on the basis of their spatial positions, we do observe that DELVE 3 lies in the broad region of sky within which satellites of the Magellanic Clouds have been discovered \citep{2015ApJ...807...50B, 2015ApJ...805..130K}. It is therefore worthwhile to consider whether DELVE 3 may also be an accreted satellite associated with the Magellanic Clouds.
\par At present, DELVE 3 is located at a (3D) separation of 43 kpc from the LMC and 32 kpc from the SMC, compared to its 49.5 kpc separation from the Galactic Center. Comparing the on-sky position and distance of DELVE 3 to the numerical simulations of \citet{2016MNRAS.461.2212J}, we found that the system resides at a position where the expected density of LMC satellite debris is much lower than many of the DES LMC satellite candidates. However, the system's Galactocentric distance falls within the range where it is plausible for a connection between the systems to exist. Plotting the solar-reflex-corrected proper motion vector for DELVE 3, we also observed that DELVE 3 is moving in the same broad direction as multiple dwarf galaxies believed to be associated with the LMC/SMC \citep{2018ApJ...867...19K,2020ApJ...893..121P}. Thus, we see no reason to rule out a possible association between DELVE 3 and the Magellanic Clouds at this time, but we emphasize that spectroscopy (and a higher-confidence proper motion measurement) will be required to confirm or dispute a connection.

\section{Summary}
\label{sec:summary}

We have presented the discovery of six ultra-faint stellar systems, all detected in the second data release of the DECam Local Volume Exploration survey. To better characterize the properties of these systems, we obtained deeper imaging of all six systems with either DECam  or GMOS-N. For the four candidates with GMOS-N imaging, the excellent depth and image quality of the observations allowed us to probe stars several magnitudes deeper than the main sequence turnoff. While the DECam imaging for the remaining two candidates represented only a minor improvement over the discovery data, we were nonetheless able to characterize the candidates' structure. Based on maximum-likelihood fits, we determined that three of the six systems are more spatially extended $(r_{1/2} > 15$ pc), while the other three systems are more compact $(r_{1/2} < 10$ pc). We measured the proper motions of five of these six candidates, although two of these measurements are tentative due to uncertainty in the membership of the stars from which they were derived.
\par The three compact systems, which we named DELVE 3, DELVE 4, and DELVE 5, likely represent ultra-faint star clusters. These systems all feature an old, metal-poor stellar population, favoring an accreted origin for each. Toward this end, we argued that DELVE 3 may represent a distant satellite of the Magellanic Clouds, and that DELVE 4 and DELVE 5 may be members of the Milky Way satellite plane known as the Vast Polar Structure (VPOS). Radial velocities will be needed to test these possibilities.
\par The three remaining candidates, which we named Virgo II, Bo\"{o}tes~V, and Leo Minor I, most likely represent ultra-faint dwarf galaxies on the basis of their sizes and luminosities. However, we could not rule out a star cluster nature for these systems, and spectroscopic velocity and/or metallicity dispersion measurements will be necessary to assess their nature. We explored whether Leo Minor I and Virgo II may be associated with the Sagittarius dwarf spheroidal, and found that the latter may possibly be associated. We also found that Bo\"{o}tes~V is a promising VPOS member candidate, potentially adding to the roster of systems associated with this planar structure.

\par Our results offer a preview of the ultra-faint satellite systems soon to be discovered by the Vera C. Rubin Observatory and its Legacy Survey of Space and Time (LSST; \citealt{2019ApJ...873..111I}). In particular, it is expected that a significant majority of the Milky Way satellite galaxies and star clusters that the Rubin LSST will soon discover will lie at luminosities similar to the candidates presented here ($M_V \gtrsim -3.2$\,mag; see e.g., \citealt{2014ApJ...795L..13H,2019ApJ...873..111I,2019ApJ...873...34N,2022MNRAS.tmp.2274M}). Although candidates in this luminosity range have repeatedly been discovered since SDSS, the
Rubin LSST will dramatically increase the volume within which we can detect these systems. As is already apparent, the physical classification of these faint systems will be one of the major challenges in the era of Rubin.

\section{Acknowledgments}
It is a pleasure to thank Simon Smith, Alan McConnachie, and the UNIONS collaboration for pleasant and productive discussions during which we coordinated the submission of our independent manuscripts reporting the discovery of Bo\"{o}tes~V. \par We thank the staff of Gemini Observatory North and Cerro Tololo Inter-American Observatory for their support in the execution of our observations, and we are grateful to the directors of each observatory for granting our requests for Director's Discretionary time to study some of the candidates presented here.
\par This project is partially supported by the NASA Fermi Guest Investigator Program Cycle 9 No.\ 91201. This work is partially supported by Fermilab LDRD project L2019-011. W.C. gratefully acknowledges support from a Gruber Science Fellowship at Yale University. A.B.P. acknowledges support from NSF grant AST-1813881. A.H.R. acknowledges support from an NSF Graduate Research Fellowship through grant DGE-1746932 and a Research Fellowship from the Royal Commission for the Exhibition of 1851. R. R. M. gratefully acknowledges support by the ANID BASAL project FB210003 and ANID Fondecyt  project 1221695.
\par This work was enabled in part by observations made from the Gemini North telescope, located within the Maunakea Science Reserve and adjacent to the summit of Maunakea. We are grateful for the privilege of observing the Universe from a place that is unique in both its astronomical quality and its cultural significance. 
\par This work is based in part on observations obtained at the international Gemini Observatory, a program of NSF’s NOIRLab (processed using DRAGONS (Data Reduction for Astronomy from Gemini Observatory North and South), which is managed by the Association of Universities for Research in Astronomy (AURA) under a cooperative agreement with the National Science Foundation on behalf of the Gemini Observatory partnership: the National Science Foundation (United States), National Research Council (Canada), Agencia Nacional de Investigaci\'{o}n y Desarrollo (Chile), Ministerio de Ciencia, Tecnolog\'{i}a e Innovaci\'{o}n (Argentina), Minist\'{e}rio da Ci\^{e}ncia, Tecnologia, Inova\c{c}\~{o}es e Comunica\c{c}\~{o}es (Brazil), and Korea Astronomy and Space Science Institute (Republic of Korea).
\par Some of the data presented herein were obtained at the W. M. Keck Observatory, which is operated as a scientific partnership among the California Institute of Technology, the University of California and the National Aeronautics and Space Administration. The Observatory was made possible by the generous financial support of the W. M. Keck Foundation.

This project used data obtained with the Dark Energy Camera (DECam), 
which was constructed by the Dark Energy Survey (DES) collaboration.
Funding for the DES Projects has been provided by 
the DOE and NSF (USA),   
MISE (Spain),   
STFC (UK), 
HEFCE (UK), 
NCSA (UIUC), 
KICP (U. Chicago), 
CCAPP (Ohio State), 
MIFPA (Texas A\&M University),  
CNPQ, 
FAPERJ, 
FINEP (Brazil), 
MINECO (Spain), 
DFG (Germany), 
and the collaborating institutions in the Dark Energy Survey, which are
Argonne Lab, 
UC Santa Cruz, 
University of Cambridge, 
CIEMAT-Madrid, 
University of Chicago, 
University College London, 
DES-Brazil Consortium, 
University of Edinburgh, 
ETH Z{\"u}rich, 
Fermilab, 
University of Illinois, 
ICE (IEEC-CSIC), 
IFAE Barcelona, 
Lawrence Berkeley Lab, 
LMU M{\"u}nchen, and the associated Excellence Cluster Universe, 
University of Michigan, 
NSF's National Optical-Infrared Astronomy Research Laboratory, 
University of Nottingham, 
Ohio State University, 
OzDES Membership Consortium
University of Pennsylvania, 
University of Portsmouth, 
SLAC National Lab, 
Stanford University, 
University of Sussex, 
and Texas A\&M University.

This work has made use of data from the European Space Agency (ESA) mission {\it Gaia} (\url{https://www.cosmos.esa.int/gaia}), processed by the {\it Gaia} Data Processing and Analysis Consortium (DPAC, \url{https://www.cosmos.esa.int/web/gaia/dpac/consortium}).
Funding for the DPAC has been provided by national institutions, in particular the institutions participating in the {\it Gaia} Multilateral Agreement.

Based on observations at Cerro Tololo Inter-American Observatory, NSF's National Optical-Infrared Astronomy Research Laboratory (2019A-0305; PI: Drlica-Wagner), which is operated by the Association of Universities for Research in Astronomy (AURA) under a cooperative agreement with the National Science Foundation.

This manuscript has been authored by Fermi Research Alliance, LLC, under contract No.\ DE-AC02-07CH11359 with the US Department of Energy, Office of Science, Office of High Energy Physics. The United States Government retains and the publisher, by accepting the article for publication, acknowledges that the United States Government retains a non-exclusive, paid-up, irrevocable, worldwide license to publish or reproduce the published form of this manuscript, or allow others to do so, for United States Government purposes.

The Legacy Surveys consist of three individual and complementary projects: the Dark Energy Camera Legacy Survey (DECaLS; Proposal ID 2014B-0404; PIs: David Schlegel and Arjun Dey), the Beijing-Arizona Sky Survey (BASS; NOAO Prop. ID 2015A-0801; PIs: Zhou Xu and Xiaohui Fan), and the Mayall z-band Legacy Survey (MzLS; Prop. ID 2016A-0453; PI: Arjun Dey). DECaLS, BASS and MzLS together include data obtained, respectively, at the Blanco telescope, Cerro Tololo Inter-American Observatory, NSF’s NOIRLab; the Bok telescope, Steward Observatory, University of Arizona; and the Mayall telescope, Kitt Peak National Observatory, NOIRLab. Pipeline processing and analyses of the data were supported by NOIRLab and the Lawrence Berkeley National Laboratory (LBNL). The Legacy Surveys project is honored to be permitted to conduct astronomical research on Iolkam Du’ag (Kitt Peak), a mountain with particular significance to the Tohono O’odham Nation.

NOIRLab is operated by the Association of Universities for Research in Astronomy (AURA) under a cooperative agreement with the National Science Foundation. LBNL is managed by the Regents of the University of California under contract to the U.S. Department of Energy.

This project used data obtained with the Dark Energy Camera (DECam), which was constructed by the Dark Energy Survey (DES) collaboration. Funding for the DES Projects has been provided by the U.S. Department of Energy, the U.S. National Science Foundation, the Ministry of Science and Education of Spain, the Science and Technology Facilities Council of the United Kingdom, the Higher Education Funding Council for England, the National Center for Supercomputing Applications at the University of Illinois at Urbana-Champaign, the Kavli Institute of Cosmological Physics at the University of Chicago, Center for Cosmology and Astro-Particle Physics at the Ohio State University, the Mitchell Institute for Fundamental Physics and Astronomy at Texas A\&M University, Financiadora de Estudos e Projetos, Fundacao Carlos Chagas Filho de Amparo, Financiadora de Estudos e Projetos, Fundacao Carlos Chagas Filho de Amparo a Pesquisa do Estado do Rio de Janeiro, Conselho Nacional de Desenvolvimento Cientifico e Tecnologico and the Ministerio da Ciencia, Tecnologia e Inovacao, the Deutsche Forschungsgemeinschaft and the Collaborating Institutions in the Dark Energy Survey. The Collaborating Institutions are Argonne National Laboratory, the University of California at Santa Cruz, the University of Cambridge, Centro de Investigaciones Energeticas, Medioambientales y Tecnologicas-Madrid, the University of Chicago, University College London, the DES-Brazil Consortium, the University of Edinburgh, the Eidgenossische Technische Hochschule (ETH) Zurich, Fermi National Accelerator Laboratory, the University of Illinois at Urbana-Champaign, the Institut de Ciencies de l’Espai (IEEC/CSIC), the Institut de Fisica d’Altes Energies, Lawrence Berkeley National Laboratory, the Ludwig Maximilians Universitat Munchen and the associated Excellence Cluster Universe, the University of Michigan, NSF’s NOIRLab, the University of Nottingham, the Ohio State University, the University of Pennsylvania, the University of Portsmouth, SLAC National Accelerator Laboratory, Stanford University, the University of Sussex, and Texas A\&M University.

BASS is a key project of the Telescope Access Program (TAP), which has been funded by the National Astronomical Observatories of China, the Chinese Academy of Sciences (the Strategic Priority Research Program “The Emergence of Cosmological Structures” Grant \#XDB09000000), and the Special Fund for Astronomy from the Ministry of Finance. The BASS is also supported by the External Cooperation Program of Chinese Academy of Sciences (Grant \#114A11KYSB20160057), and Chinese National Natural Science Foundation (Grant \#12120101003, \#11433005).

The Legacy Survey team makes use of data products from the Near-Earth Object Wide-field Infrared Survey Explorer (NEOWISE), which is a project of the Jet Propulsion Laboratory/California Institute of Technology. NEOWISE is funded by the National Aeronautics and Space Administration.

The Legacy Surveys imaging of the DESI footprint is supported by the Director, Office of Science, Office of High Energy Physics of the U.S. Department of Energy under Contract No. DE-AC02-05CH1123, by the National Energy Research Scientific Computing Center, a DOE Office of Science User Facility under the same contract; and by the U.S. National Science Foundation, Division of Astronomical Sciences under Contract No. AST-0950945 to NOAO.

\facility{Blanco (DECam), Gemini:Gillett (GMOS-N), \Gaia, Astro Data Lab, Astro Data Archive.}
\software{\code{astropy} \citep{astropy:2013,astropy:2018}, \emcee \citep{Foreman-Mackey:2013}, \code{fitsio},\footnote{\url{https://github.com/esheldon/fitsio}} \healpix \citep{Gorski:2005},\footnote{\url{http://healpix.sourceforge.net}} \code{healpy},\footnote{\url{https://github.com/healpy/healpy}} \code{Matplotlib} \citep{Hunter:2007}, \code{numpy} \citep{numpy:2011}, \code{scipy} \citep{scipy:2001}, \ugali \citep{2015ApJ...807...50B}\footnote{\url{https://github.com/DarkEnergySurvey/ugali}}, \code{gala} \citep{gala}.}

\bibliography{main}

\appendix
\FloatBarrier
\section{Completeness Curves for the GMOS-N Photometry}
\begin{figure*}[h]

    \centering
    \includegraphics[width = \textwidth]{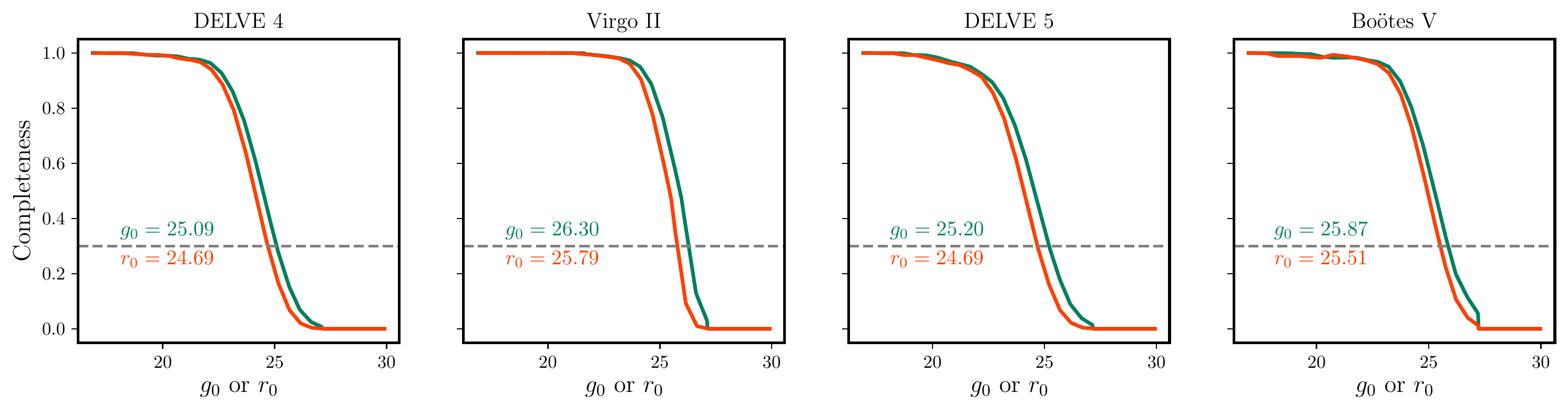}
    \caption{Completeness as a function of extinction-corrected magnitude ($g_0$ or $r_0$) for the four candidates with GMOS-N imaging, after application of the $\chi$ and \textit{sharp} cuts described in the main text. The $g$-band completeness curve is shown in green, while the $r$-band completeness curve is shown in red. The 30\% completeness limit for both bands is denoted by a gray dashed line, with the corresponding magnitude limits labeled above and below; these match the values given in \tabref{obsTable}. }
    \label{fig:completeness}
\end{figure*}

In \figref{completeness}, we depict the completeness of our GMOS-N photometry as a function of magnitude, derived from the artificial star tests described in \secref{gmosphotometry}.
\vspace{-13em}
\label{app:complete}

\pagebreak
\section{Candidate Members Stars Identified based on \Gaia Proper Motions}
\label{app:appPMs}
In \tabref{PMs}, we list the member stars in each candidate that have proper motions measured by \Gaia. We report only stars with mixture model probabilities $p_{MM}>0.01$ (see \secref{propermotions} for details). The one exception to this probability cut is Virgo~II, for which we report the two stars that may individually be members but that have discrepant proper motions (see \secref{virgoII}).
\begin{deluxetable}{l c c c c}[H]
\tablecolumns{5}
\tablewidth{0pt}
\tabletypesize{\small}
\tablecaption{\label{tab:PMs} Candidate Proper-Motion Members for Each Host Reported in this Work}
\tablehead{\colhead{Host} &  \colhead{Source ID}  &  \colhead{RA (deg)}  & \colhead{Dec. (deg)} &  \colhead{$p_{\rm MM}$} }
\startdata
Virgo II & 1159816202723001344 & 225.055304 & 5.909133 & 0.0\\
Virgo II & 1159816404586450304 & 225.044828 & 5.916107 & 0.0 \\
\hline
Bo\"{o}tes~V & 1478121378694742912* & 213.909901 & 32.906644 & 1.00 \\
Bo\"{o}tes~V & 1478121382994560256* & 213.911812 & 32.914049 & 1.00 \\
Bo\"{o}tes~V & 1478121413054482560 & 213.921037 & 32.922418 & 1.00 \\
Bo\"{o}tes~V & 1478121584853173888 & 213.907406 & 32.912497 & 1.00 \\
Bo\"{o}tes~V & 1478121584853174272* & 213.902565 & 32.919849 & 1.00 \\
Bo\"{o}tes~V & 1478121722292128768* & 213.930665 & 32.939983 & 0.99 \\
Bo\"{o}tes~V & 1478126743113866624 & 214.002746 & 32.945834 & 0.04 \\
Bo\"{o}tes~V & 1478127773905897344 & 213.974961 & 32.982115 & 0.06 \\
\hline
Leo~Minor~I & 731478564536994432 & 164.272010 & 28.854826 & 0.99 \\
Leo~Minor~I & 731478663321185152 & 164.274357 & 28.876532 & 0.98 \\
Leo~Minor~I & 732979913304939776 & 164.285114 & 28.893772 & 0.96 \\
\hline
DELVE~3 & 6445895198535725568 & 290.409730 & -60.779749 & 0.98 \\
DELVE~3 & 6445895232895460096 & 290.374884 & -60.784377 & 0.99 \\
DELVE~3 & 6445895645212360192 & 290.407766 & -60.747341 & 0.05 \\
\hline
DELVE~4 & 1270955387816956928* & 230.761381 & 27.359978 & 1.00 \\
DELVE~4 & 1270955387817087232* & 230.768366 & 27.367744 & 0.99 \\
DELVE~4 & 1270955800133818240* & 230.779884 & 27.384258 & 1.00 \\
DELVE~4 & 1270956006292249984 & 230.774651 & 27.394922 & 1.00 \\
DELVE~4 & 1270956006292370176 & 230.785858 & 27.394184 & 1.00 \\
DELVE~4 & 1270956006292379648* & 230.773749 & 27.395964 & 1.00 \\
\hline
DELVE~5 & 1236269365075548672 & 222.081767 & 17.409420 & 0.79 \\
DELVE~5 & 1236270945623558912 & 222.094775 & 17.473537 & 1.00 \\
DELVE~5 & 1236271564098855296 & 222.178859 & 17.481047 & 0.03 \\
DELVE~5 & 1236317949745662336 & 222.075186 & 17.496322 & 0.92 \\
\enddata
\tablecomments{Stars with an asterisk refer to those that overlap with the candidate member stars identified by the \Gaia-based wavelet transform search algorithm from \citet{2021ApJ...915...48D}.}
\end{deluxetable}
\FloatBarrier
\section{Testing the Robustness of the GMOS-N Structural Fits}

\subsection{Injection-Recovery Tests with Simulated Stellar Systems}
\label{app:simtests}
\begin{figure*}[t]
    \centering
    \includegraphics[width=0.95\textwidth]{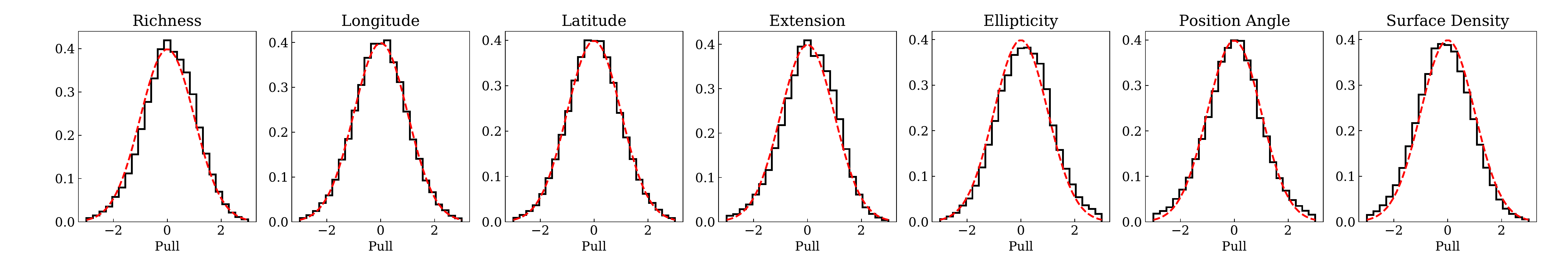}
    \caption{Pull distribution for each structural parameter in the binned Poisson maximum-likelihood fit of the GMOS-N data. The pull is calculated as the difference between the fit value of the parameter and the input parameter divided by the fit uncertainty. The black histogram shows the pull distribution derived from $\roughly45,000$ simulations, while the red dashed line shows a standard unit normal distribution, $\mathcal{N}(0,1)$.}
    \label{fig:sims_pull}
\end{figure*}
\begin{deluxetable}{c c c}
\tablecolumns{3}
\tablewidth{0pt}
\tabletypesize{\small}
\tablecaption{\label{tab:sims} Range of simulated parameters}
\tablehead{\colhead{Parameter} &  \colhead{Range}  &  \colhead{Unit}}
\startdata
Observed Stars (Richness) & [25, 250] & \\
Centroid Coordinate ($x$) & [1100, 2200] & pixels \\
Centroid Coordinate ($y$) & [500, 1750] & pixels \\
Extension ($a_h$) & [50, 600] & pixels \\
Ellipticity ($\epsilon$) & [0, 0.8] & \\
Position Angle (PA) & [0, 180] & deg \\
Surface Density ($\Sigma$) & [1, 50] & stars/Mpix
\enddata
\end{deluxetable}

To validate the performance of our binned Poisson maximum-likelihood structural fit of the GMOS-N data, we performed a suite of simulation-injection-recovery tests.
We generated simulated satellites over a range of parameters that span our observed candidates (\tabref{sims}).
For each choice of input parameters, we simulated the satellite and foreground stellar distribution by drawing a Poisson random sample of stars from our model with the detection fraction mask applied.
We fit the simulated stellar distribution using the same procedure that we applied to the GMOS-N data, characterizing posterior probability distribution with the Metropolis-Hastings sampler, \code{emcee}.
We estimated the best-fit parameters and their uncertainties by calculating the median, 16th percentile ($p_{16}$), and the 84th percentile ($p_{84}$) of the posterior probability distribution.
To assess the ability of our algorithm to recover the input parameter values within the derived uncertainties, we calculated the pull distribution, $\mathcal{P}$, for each parameter using the asymmetric definition from \citet{Demortier:2002},

\begin{equation}
  \mathcal{P}(\theta) = \begin{cases}
    \dfrac{(\theta - \theta_0)}{\theta_{\rm UL} - \theta}, & \text{if $\theta \leq \theta_0$}. \\[+1em]
    \dfrac{(\theta - \theta_0)}{\theta_{\rm LL} - \theta}, & \text{otherwise},
  \end{cases}
\end{equation}

\noindent where $\theta$ is the best-fit value of the parameter (i.e., the median of the posterior), $\theta_{\rm UL}$ and $\theta_{\rm LL}$ are respectively the upper and lower limit on the best-fit value (i.e., estimated from the minimum interval containing the peak and $68\%$ of the posterior), and $\theta_0$ is the input value of the parameter. 
In the case that our procedure accurately captures the statistical scatter in the fit parameter and its uncertainty, the pull distribution will be described by a unit normal distribution, $\mathcal{P}(\theta) \sim \mathcal{N}(0, 1)$.
The pull distributions for the set of input parameters over our full suite of simulations can be seen in \figref{sims_pull}.
To explore the dependence of our fits on the input parameter values, we also examined the mean and standard deviation of the pull distribution binned by input parameters.
\par We find that the pull distribution is well-described by $\mathcal{N}(0,1)$ over the sampled parameter space.
However, there are several regions where our peak/interval summary statistics are inadequate to describe the posterior distribution.
This generally occurs when the posterior runs up against bounds on the allowed ranges that parameters can take. 
We describe specific cases below.

\begin{enumerate}
    \item At small values of ellipticity, the posterior is bounded by the prior of $\epsilon \geq 0$. In these cases, the posterior is one-sided, peaking at $\epsilon = 0$. Our peak/interval summary statistics are insufficient to describe the posterior distribution, and our pull distribution will be biased. This is not particularly concerning because for real systems in this regime, we instead quote an upper limit on $\epsilon$ derived from the 95th percentile of the posterior distribution. 

    \item In cases where the $\epsilon \sim 0$, the position angle is poorly constrained. However, because the position angle is bounded to values $0 < {\rm PA} < 180$ deg, there will be a maximum separation between the fit and true values, as well as a maximum size of the 68\% interval. Due to these bounds, the pull distribution is not well summarized by a unit normal distribution.

    \item The richness is restricted to positive values. In cases where the true richness is low (richness $\sim 25$) and the extension is large (extension $> 400$ pix), the posterior probability distribution for the richness will sometimes run up against the bound at richness = 0. Similar to the previous two cases, these situations would be identified in real systems when the posterior was inspected. In these cases, it is likely that the system would have been rejected from the sample.
\end{enumerate}

Each of these situations would be identified during inspection of the candidate fit results, and a different technique would be applied to summarize the posterior probability distribution (e.g., deriving a 84\% confidence level upper limit). Regardless, these cases are included in the simulations shown in \figref{sims_pull}, and it can be seen that they have a relatively small impact in overall performance of the algorithm.

\end{document}